\documentclass[%
superscriptaddress,
reprint,
 amsmath,amssymb,
 aps,
pra,
nofootinbib
]{revtex4-1}

\usepackage{graphicx}
\usepackage[percent]{overpic}
\usepackage{dcolumn}
\usepackage{bm}

\usepackage{hyperref}
\usepackage{color} 
\definecolor{rltred}{rgb}{0.75,0,0}
\definecolor{rltgreen}{rgb}{0,0.6,0}
\definecolor{rltblue}{rgb}{0.3,0.3,1}

\usepackage{placeins}

\hypersetup{colorlinks,%
    hypertexnames=true,%
    linkcolor=rltred,%
    citecolor=rltgreen,%
    linktocpage=true}

\begin{document}
\title{Propagating two-particle reduced density matrices without wavefunctions}

\author{Fabian Lackner}
\email{fabian.lackner@tuwien.ac.at}
\affiliation{Institute for Theoretical Physics, Vienna University of Technology,
Wiedner Hauptstra\ss e 8-10/136, 1040 Vienna, Austria, EU} 

\author{Iva B\v rezinov\'a}
\email{iva.brezinova@tuwien.ac.at}
\affiliation{Institute for Theoretical Physics, Vienna University of Technology,
Wiedner Hauptstra\ss e 8-10/136, 1040 Vienna, Austria, EU} 

\author{Takeshi Sato}
\affiliation{Photon Science Center, Graduate School of Engineering, The University of Tokyo,7-3-1 Hongo, Bunkyo-ku, Tokyo 113-8656, Japan} 

\author{Kenichi L. Ishikawa}
\affiliation{Photon Science Center, Graduate School of Engineering, The University of Tokyo,7-3-1 Hongo, Bunkyo-ku, Tokyo 113-8656, Japan} 
\affiliation{Department of Nuclear Engineering and Management, School of Engineering, The University of Tokyo, 7-3-1 Hongo, Bunkyo-ku, Tokyo 113-8656, Japan} 

\author{Joachim Burgd\"orfer}
\affiliation{Institute for Theoretical Physics, Vienna University of Technology,
Wiedner Hauptstra\ss e 8-10/136, 1040 Vienna, Austria, EU}  

\date{\today}

\begin{abstract}
Describing time-dependent many-body systems where correlation effects play an important role remains a major theoretical challenge. In this paper we develop a time-dependent many-body theory that is based on the two-particle reduced density matrix (2-RDM). We develop a closed equation of motion for the 2-RDM employing a novel reconstruction functional for the three-particle reduced density matrix (3-RDM) that preserves norm, energy, and spin symmetries during time propagation. We show that approximately enforcing $N$-representability during time evolution is essential for achieving stable solutions. As a prototypical test case which features long-range Coulomb interactions we employ the one-dimensional model for lithium hydride (LiH) in strong infrared laser fields. We probe both one-particle observables such as the time-dependent dipole moment and two-particle observables such as the pair density and mean electron-electron interaction energy. Our results are in very good agreement with numerically exact solutions for the $N$-electron wavefunction obtained from the multiconfigurational time-dependent Hartree-Fock method.
\end{abstract}
\pacs{} 
\maketitle

\section{\label{sec:int}Introduction}
The direct solution of the time-dependent $N$-particle Schr\"odinger equation has remained a major challenge for systems with a large number of particles $N$. This is in particular true for the time-dependent many-electron problem in atoms, molecules, and condensed matter with the long-range Coulomb interactions and Coulomb continua ubiquitously present. Numerically exact solutions have become available only for small systems such as He \cite{FeiNagPaz08,PazFeiNag12,HocBon11,ParSmyTay98,PalResMcC08} or H$_2$ \cite{SanPalCar07,PalBacMar07,GuaBarSch10,Sae00,VanHorMar06,DehBanKam10} as the numerical effort grows factorially with particle number $N$. Ground state properties of large systems involving tens to hundreds of particles can routinely be calculated employing sophisticated methods developed in quantum chemistry and solid state physics such as configuration interaction methods, coupled cluster methods, perturbative methods, and density functional theory (DFT) (see e.g.~\cite{SzaOst89,ParYan89}). An analogous development for time-dependent systems and systems far from the ground state is still in its infancy. The time-dependent extension of DFT, the time-dependent density functional theory (TDDFT) (for a review see \cite{Ull12}) features a favorable linear scaling with $N$ and allows the approximate treatment of large and extended systems (see e.g.~\cite{wac14,bau01,oto08,dri11,bur05,poh00}). However, accurate exchange-correlation functionals beyond the adiabatic limit containing memory effects are not yet known. Alternatively, the so-called time-dependent current-density functional theory (TDCDFT) has been proposed (for a review see \cite{Ull12}) for which, up to now, however only few approximations for the exchange-correlation vector potential have become available \cite{VigKoh96}. On a more conceptual level, only physical observables that are explicit functionals of the reduced one-particle density (or current density) can be easily determined from TDDFT. Read-out functionals of two-particle observables are still largely missing \cite{LapLee98,WilBau07,RohSimBur06}.\\
Extensions of the direct solution of the $N$-electron Schr\"odinger equation beyond the two-electron problem employs the multiconfigurational time-dependent Hartree-Fock method (MCTDHF) (\cite{CaiZanKit05,hochstuhl_time-dependent_2014}). In principle, the MCTDHF method converges to the numerically exact solution if a sufficient number of orbitals and configurations is used. However, its numerical effort scales factorially with the number of particles. A recently proposed variant, the time-dependent complete active space self-consistent field (TD-CASSCF) method \cite{TakKen13} which, in analogy to its ground state counterpart, decomposes the state space into frozen, dynamically polarizable, and dynamically active orbitals can considerably reduce the numerical effort yet eventually still leads to an factorial scaling with the number of active electrons $N^\star$ ($N^\star<N$). \\
Our point of departure is the recent advance in the ground-state description of larger electronic systems employing the two-particle reduced density matrix (2-RDM). Going back to the pioneering work in the 1950s \cite{Low55,Bop59}, the 2-RDM method has recently matured to accuracies that often outperform those of CCSD(T) at similar or smaller numerical cost (see e.g.~\cite{Maz04,ZhaBraOve03,Maz06_antiher}). Similar to DFT, this method bypasses the need for the $N$-particle wavefunction but employs the 2-RDM rather than the one-particle density as the fundamental quantity. Unlike DFT, however, the energy and all two-particle observables can be expressed exactly in terms of the 2-RDM without invoking an approximate exchange-correlation functional or read-out functional. Proposed methods for calculating the 2-RDM include variational minimization of the energy as a functional of the 2-RDM, solution of the contracted Schr\"odigner equation, and solution of the antihermitian part of the contracted Schr\"odinger equation (for a review see \cite{Maz07}). A major challenge in applying the 2-RDM method is to enforce $N$-representability conditions, i.e.~to constrain the trial 2-RDMs to those that represent reductions of either pure or ensembles of fermionic $N$-particle states \cite{Col63,GarPer64,Erd78}. Despite recent progress \cite{Maz12}, a complete list of (pure state) $N$-representability conditions is not known and one is limited to few necessary but not sufficient $N$-representability conditions in numerical implementations.\\
The present work aims at extending this theoretical development to the time-dependent 2-RDM (TD-2RDM) in the presence of external time-dependent potentials. A prototypical case is an $N$-electron system driven by a (moderately) strong laser field. The ultimate goal is to propagate 2-RDMs without invoking wavefunctions. There have been only few previous attempts along these lines for nuclear \cite{GheKriRei93,SchReiToe90,TohSch14}, atomic \cite{BunNes08} and condensed matter systems \cite{AkbHasRub12}. They all encountered instabilities due to the intrinsic nonlinearity of the equation of motion for the 2-RDM resulting in the violation of positive definiteness.\\
In the present paper we take two major steps towards an accurate TD-2RDM method which scales polynomially with particle number. We develop a novel reconstruction functional that allows closure of the equation of motion for the 2-RDM without introducing uncontrolled violations of norm, spin, and energy conservation. Secondly, we impose two necessary $N$-representability constraints ``on the fly" during the time evolution thereby controlling dynamical instabilities previously observed \cite{GheKriRei93,SchReiToe90,TohSch14,BunNes08,AkbHasRub12}. As a prototypical test case we apply our method to the electronic dynamics of a 1D model of LiH (a four electron system) in strong laser fields of up to $I=8\times10^{14}{\rm W/cm^2}$. We investigate both the linear as well as nonlinear response of this system. For this system, numerically exact results can be determined by the MCTDHF method against which we gauge our results. This method also serves as a source for the initial state within the TD-2RDM calculations for which we take the field-free ground state. We compare with results from TDDFT and time-dependent Hartree-Fock (TDHF) calculations.\\
The paper is structured as follows: In Sec.~\ref{sec:equ_2rdm} we briefly review the equation of motion for the 2-RDM being one element of the Bogoliubov-Born-Green-Kirkwood-Yvon (BBGKY) hierarchy and its closure in terms of approximate generalized collision integrals. We introduce a new reconstruction functional for these integrals in Sec.~\ref{sec:con}. Preserving a stable dynamical evolution requires enforcing $N$-representability constraints (or ``purification") ``on the fly", implementation of which is discussed in Sec.~\ref{sec:N-r}. Numerical results for typical one-body observables such as the time-dependent dipole moment and two-body observables (interaction energy) for LiH are presented in Sec.~\ref{sec:orb}. Throughout this paper we use atomic units ($e=\hbar=m=1$).
\section{\label{sec:equ_2rdm} Equation of motion for the 2-RDM}

\subsection{\label{sec:basic}Basic properties}
The $p$-particle reduced density matrix ($p$-RDM) $D(x_1 \dots x_p;x'_1 \dots x'_p;t)$ of an $N$-particle system in a pure state $\Psi$ is determined by tracing out the coordinates of the remaining  $N-p$ particles from the bilinear form $\Psi\Psi^*$,
\begin{align}\label{eq:prdm}
&D(x_1,\dots,x_p;x'_1,\dots, x'_p;t)=  \nonumber\\ 
&\frac{N!}{(N-p)!} \int \Psi(x_1\dots x_p,x_{p+1} \dots x_N,t) \nonumber \\
&\times \Psi^*(x'_1\dots x'_p,x_{p+1} \dots x_N,t) \text{d}x_{p+1}...\text{d}x_N,
\end{align}
where $x_i=(z_i,\sigma_i)$ comprises the space coordinate $z_i$ and the spin coordinate $\sigma_i \in \{\uparrow,\downarrow\}$. The $p$-RDMs are hermitian and antisymmetric with respect to exchange of the $x$ or $x'$ variables. We normalize the $p$-RDMs to $\frac{N!}{(N-p)!}$. Following \cite{Bon98}, we use for the 2-RDM (Eq.~\ref{eq:prdm} with $p=2$) the following short-hand notation
\begin{align}\label{eq:2rdm_short}
	D_{12}(t) = D(x_1,x_2;x'_1,x'_2;t).
\end{align}
The equation of motion for $D_{12}$ is given by the second member of the BBGKY hierarchy as \cite{Bon98} 
\begin{align}\label{eq:eom_d12}
	i\partial_tD_{12} &= \left[H_{12},D_{12}\right] + {\rm Tr}_3\left[W_{13}+W_{23},D_{123}\right] \nonumber \\
	&= \left[H_{12},D_{12}\right] + C_{12}\left[D_{123}\right],
\end{align}
where $D_{123}$ is the 3-RDM and $H_{12}$ denotes the two-particle Hamiltonian
\begin{align}\label{eq:2p_ham}
	H_{12} =h_1+h_2+W_{12},
\end{align}
with $h_i$ the one-particle part containing the kinetic energy and the explicitly time-dependent external field, and $W_{12}$ the electron-electron interaction. For the specific example of the Hamiltonian of the 1D LiH molecule see Eq.~\ref{eq:hamiltonian} below. \\
Equation~\ref{eq:eom_d12} is not closed but depends on the next higher-order RDM through the three-body collision operator 
\begin{align}\label{eq:C12}
	C_{12}[D_{123}] ={\rm Tr}_3\left[W_{13}+W_{23},D_{123}\right],
\end{align}
where the partial trace extends over the third particle of the commutator between the interaction potential and the 3-RDM. Approximately solving Eq.~\ref{eq:eom_d12} thus inevitably requires closure, i.e.~approximating $D_{123}$ and the resulting collision operator $C_{12}$ by quantities already determined by evolution of the 2-RDM, i.e.~$D_1$ and $D_{12}$. The following quantities
\begin{align}\label{eq:r123}
	D^{\rm R}_{123}[D_{12}]\approx D_{123},
\end{align}
and 
\begin{align}\label{eq:approx_c}
	C^{R}_{12}[D_{12}] = C_{12}[D^{\rm R}_{123}[D_{12}]] ,
\end{align}
are referred to as the reconstruction functional for the 3-RDM ($D^{\rm R}_{123}$), and the collision operator ($C^{\rm R}_{12}$), respectively. Employing such a reconstruction the equation of motion for the 2-RDM has the closed form
\begin{align}\label{eq:eom_approx}
 i\partial_tD_{12} = \left[H_{12},D_{12}\right] + C^{\rm R}_{12}\left[D_{12}\right].
\end{align}
Equation~\ref{eq:eom_approx} must conserve invariants of the $N$-particle system. These include the norm (or particle number), the energy (for time-independent Hamilton operators), and spin (for spin-independent interactions). Some of these conservation laws provide constraints on admissible reconstruction functionals. Conservation of particle number follows immediately from 
\begin{align}\label{eq:norm}
	i\partial_t{\rm Tr}_{12}D_{12} &= {\rm Tr}_{12}\left[H_{12},D_{12}\right] \nonumber \\
	&+{\rm Tr_{123}}\left[W_{13}+W_{23},D^{\rm R}_{123}\right]=0
\end{align}
using the permutation symmetry of traces and the antisymmetry of the commutator. Equation~\ref{eq:norm}, thus, does not provide any constraints on the reconstructed 3-RDM $D^{\rm R}_{123}$.\\
Time evolution of the energy,
\begin{align}\label{eq:energy}
E(t) =\frac{1}{2}{\rm Tr}_{12}\left(\tilde{H}_{12}D_{12}\right)
\end{align}
with
\begin{align}\label{eq:ham_energy}
\tilde{H}_{12}=\frac{h_1+h_2}{N-1} + W_{12}
\end{align}
is described by the differential equation
\begin{align}\label{eq:energy_2}
& i\partial_tE(t)=\frac{1}{2}{\rm Tr}_{12}\Big(\big(i\partial_t\tilde{H}_{12}\big)D_{12}\Big)\nonumber \\
&+\frac{N-2}{2(N-1)}{\rm Tr_{12}}\Big([W_{12},h_1+h_2]
\big(D_{12}-\frac{1}{N-2}{\rm Tr}_3D^{\rm R}_{123}\big)\Big).
\end{align}
In the absence of time-dependent external fields (i.e.~$\partial_t\tilde{H}_{12}=0$) energy should be conserved. This condition is fulfilled if 
\begin{align}\label{eq:cond_energy}
D_{12} = \frac{1}{N-2}{\rm Tr}_3D^{\rm R}_{123}.	
\end{align}
Equation~\ref{eq:cond_energy} holds, by definition, for the exact $D_{123}$. However, it provides a constraint on the reconstructed 3-RDM, $D^{\rm R}_{123}$, that has to be fulfilled at each time step. \\
The time evolution of the 1-RDM follows from Eq.~\ref{eq:eom_approx} as
\begin{align}\label{eq:eom_d1}
i\partial_t{\rm Tr}_2D_{12} = {\rm Tr}_2[H_{12},D_{12}]+{\rm Tr}_{23}\left[W_{13}+W_{23},D^{\rm R}_{123}\right]\nonumber \\
=(N-1)[h_1,D_1]+{\rm Tr_2}\left[W_{12},D_{12}+{\rm Tr}_3D^{\rm R}_{123}\right],
\end{align}
where we have used the interrelation between the 2-RDM and the 1-RDM
\begin{align}\label{eq:interrelation}
D_1=\frac{1}{N-1}\text{Tr}_2D_{12}.
\end{align}
Equation~\ref{eq:eom_d1} reduces to the correct equation of motion for $D_1$,
\begin{align}\label{eq:eom_d1_2}
i\partial_tD_1=[h_1,D_1]+{\rm Tr}_2\left[W_{12},D_{12}\right],
\end{align}
provided the constraint Eq.~\ref{eq:cond_energy} is fulfilled. Additional constraints on the 3-RDM follow from spin conservation (see Sec.~\ref{sec:spin}). \\
Compared to wavefunction based methods which scale factorially with the number of particles the computational cost of the TD-2RDM method is independent of the particle number $N$ and depends only on the total number of basis functions. The most time consuming operation within Eq.~\ref{eq:eom_approx} is the evaluation of the collision operator (Eq.~\ref{eq:C12}) where a partial trace over the interaction potential and the 3-RDM has to be evaluated. This calculation scales as $\mathcal{O}(r^5)$ with the number of basis functions $r$ if the interaction potential is diagonal in the basis (as, e.g., in spatial representation), or as $\mathcal{O}(r^7)$ for the expansion in spin orbitals. Such basis expansions are inevitable for the efficient numerical propagation of the 2-RDM in continuous systems. 
\subsection{\label{sec:def}Orbital expansion}
The expansion of the 2-RDM in terms of $2r$ orthogonal spin orbitals facilitates not only the efficient numerical propagation of the 2-RDM and the usage of quantum chemistry codes for calculating the initial 2-RDM, it also allows to conveniently impose constraints due to spin conservation. Accordingly, we expand 
\begin{align}\label{eq:exp_D_2}
&D(x_1 x_2;x'_1 x'_2;t)=\nonumber\\
&\sum_{i_1,i_2,j_1,j_2}D_{j_1 j_2}^{i_1 i_2}(t)\phi_{i_1}(x_1,t) \phi_{i_2}(x_2,t)\phi^*_{j_1}(x'_1,t)\phi^*_{j_2}(x'_2,t),
\end{align}
with spin orbitals $\phi_{i\sigma}(x,t)=\phi_i(z,t) \otimes \vert \sigma \rangle$, where we merge the spin $\sigma \in \{\uparrow, \downarrow\}$ and orbital indices $i \in \{1 \dots r\}$. For simplicity, we drop here and in the following the labels for the $p$-RDM when using the spin-orbital representation, i.e.~$D^{i_1i_2}_{j_1j_2}=[D_{12}]^{i_1i_2}_{j_1j_2}$, $D^{i_1i_2i_3}_{j_1j_2j_3}=[D_{123}]^{i_1i_2i_3}_{j_1j_2j_3}$, since the corresponding order is already uniquely characterized by the orbital-index set.
Within second quantization these 2-RDM coefficients can be expressed as matrix elements
\begin{align}\label{eq:secondQ_D_2}
D_{j_1 j_2}^{i_1 i_2}=\langle \Psi \vert \hat{a}^{\dagger}_{i_1}\hat{a}^{\dagger}_{i_2}\hat{a}_{j_2}\hat{a}_{j_1} \vert \Psi \rangle.
\end{align}   
Generalization of Eq.~\ref{eq:secondQ_D_2} to arbitrary $p$-RDMs is obvious. In this representation the $N$-particle Hamiltonian is given by 
\begin{align}\label{eq:second_H}
\hat{H}=\frac{1}{2} \sum_{i_1,i_2,j_1,j_2}\left(\frac{h^{j_1}_{i_1}\delta^{j_2}_{i_2}+\delta^{j_1}_{i_1}h^{j_2}_{i_2}}{N-1}+W^{j_1j_2}_{i_1i_2}\right)\hat{a}^{\dagger}_{i_1}\hat{a}^{\dagger}_{i_2}\hat{a}_{j_2}\hat{a}_{j_1}, 
\end{align}
where $h^{j_1}_{i_1}$ and $W^{j_1j_2}_{i_1i_2}$ are the one- and two-electron Hamilton matrix elements in the spin-orbital basis
\begin{align}\label{}
h^{j_1}_{i_1}=\langle \phi_{j_1} \vert  h_1 \vert \phi_{i_1} \rangle,
\end{align}
\begin{align}\label{}
W^{j_1 j_2}_{i_1 i_2}=\langle \phi_{j_1} \phi_{j_2} \vert  W_{12} \vert \phi_{i_1} \phi_{i_2} \rangle.
\end{align}
The time derivative of the 2-RDM expansion coefficients contains two terms
\begin{align}\label{eq:start_dev}
i \partial_t D_{j_1 j_2}^{i_1 i_2}=&\langle \Psi \vert [ \hat{a}^{\dagger}_{i_1}\hat{a}^{\dagger}_{i_2}\hat{a}_{j_2}\hat{a}_{j_1}, \hat{H}] \vert \Psi \rangle\nonumber\\
+&\langle \Psi \vert i \partial_t(\hat{a}^{\dagger}_{i_1}\hat{a}^{\dagger}_{i_2}\hat{a}_{j_2}\hat{a}_{j_1}) \vert \Psi \rangle.
\end{align}
The second term appears only for time-dependent basis sets and can be removed if the dynamics of the orbitals satisfies
\begin{align}
\langle \phi_i\vert  \partial_t \vert \phi_j \rangle=0,
\end{align} 
which is the case for the orbital equations of motion employed here (see Eq.~\ref{eq:dphi_MCTDH} below). Using the anticommutation relation of creation and annihilation operators one obtains the spin-orbital representation of the equation of motion for the 2-RDM (Eq.~\ref{eq:eom_d12})
\begin{align}\label{eq:2p_eom_orbital}
i \partial_t D^{i_1 i_2}_{j_1 j_2}=\sum_{k_1,k_2}\big( H^{k_1 k_2}_{j_1 j_2}D^{i_1 i_2}_{k_1 k_2}-D^{k_1 k_2}_{j_1 j_2}H^{i_1 i_2}_{k_1 k_2}\big)+C^{i_1 i_2}_{j_1 j_2},
\end{align}
with
\begin{align}
&H_{j_1 j_2}^{i_1 i_2}=h^{i_1}_{j_1}\delta^{i_2}_{j_2}+\delta^{i_1}_{j_1}h^{i_2}_{j_2}
+W^{i_1 i_2}_{j_1 j_2} \\
&C_{j_1 j_2}^{i_1 i_2}=I_{j_1 j_2}^{i_1 i_2}+I_{j_2 j_1}^{i_2 i_1}-(I_{i_1 i_2}^{j_1 j_2}+I_{i_2 i_1}^{j_2 j_1})^*,
\label{eq:def_Cop}
\end{align}
and
\begin{align}
I_{j_1 j_2}^{i_1 i_2}=\sum_{k_1,k_2,k_3}W_{j_1 k_1}^{k_2 k_3}D_{k_2 j_2 k_3}^{i_1 i_2 k_1}.
\label{eq:eval_I}
\end{align}
A spin-orbital basis is a convenient computational starting point for the propagation. One choice, in the spirit of time dependent configuration interaction (TDCI) calculations, would be to treat the orbitals to be time-independent and propagate only the expansion coefficients. However, time independent orbitals will, in general, require a large number of basis orbitals to properly account for the dynamics of the system. This calls for a self-consistent optimization of the orbitals as implemented within the MCTDHF-approach \cite{hochstuhl_time-dependent_2014}. To this end we adopt the orbital equations of motion from MCTDHF:
\begin{align} \label{eq:dphi_MCTDH}
i \partial_t \phi_i(z,t) &= \hat{Q} \left({h(z)\phi_i(z,t)+\sum_{u} \hat{\Gamma}_u(z,t)[D^{-1}]^u_i} \right),
\end{align}
where 
\begin{align}
\hat{Q}=1-\sum_{i=1}^{2r} \vert \phi_i \rangle \langle \phi_i \vert 
\end{align}
is the orbital projection operator assuring unitary time evolution of the basis orbitals, $[D^{-1}]^u_i$ is the inverse of the 1-RDM in the orbital representation, and 
\begin{align}
\hat{\Gamma}_u(z,t)=\sum_{vwt} D_{u\,t}^{v\,w}\phi_v(z,t)\int \phi_w(z',t)\phi^*_t(z',t)W_{\text{12}}(z,z')\text{d}z'
\end{align}
originates from electron-electron interactions. It is this term which couples the time evolution of the orbitals to the time-evolution of the 2-RDM.
\subsection{\label{sec:spin}Spin conservation}
Since the non-relativistic Hamiltonian for atoms and molecules is spin independent, i.e.~$[H_{12},S^2]=[H_{12},S_z]=0$, with
\begin{align}
&S_z=\frac{1}{2}\sum_{i} (a^{\dagger}_{i\uparrow}a_{i\uparrow}-a^{\dagger}_{i\downarrow}a_{i\downarrow})\\
&S^2=S_z^2+S_z+S_-S_+,
\end{align}
and
\begin{align}
&S_+=\sum_{i} a^{\dagger}_{i\uparrow}a_{i\downarrow}\quad \text{and}\quad S_-=\sum_{i} a^{\dagger}_{i\downarrow}a_{i\uparrow},
\end{align}
the ground state (initial state) of the system is an eigenstate of both $S_z$ and $S^2$ and remains in this spin-state during time evolution of $\Psi(t)$ for spin-independent interactions, e.g., in the present case of a laser field in dipole approximation. In particular, for closed-shell systems with an equal number of electrons in spin up $N_{\uparrow}=N/2$ and spin down $N_{\downarrow}=N/2$ the wavefunction satisfies 
\begin{align}
S_z&|\Psi(t) \rangle =0 \label{eq:spin_Sz}\\
S_+&|\Psi(t) \rangle =0 \label{eq:spin_S2},
\end{align}
where Eq.~\ref{eq:spin_S2} together with Eq.~\ref{eq:spin_Sz} is equivalent to $S^2|\Psi(t) \rangle =0$.
These spin symmetries enforce specific symmetries on the 2-RDM that must be conserved during time propagation. \\
The most obvious symmetry originating from Eq.~\ref{eq:spin_Sz} is that the 2-RDM contains only two independent non-vanishing blocks given by
\begin{align}\label{eq:vanish}
D^{i_1 \uparrow i_2 \uparrow}_{j_1 \uparrow j_2 \uparrow} \quad \text{and} \quad D^{i_1 \uparrow i_2 \downarrow}_{j_1 \uparrow j_2 \downarrow},
\end{align}
with $i,j \in \{1\dots r \}$ for the spatial part and $\{\uparrow,\downarrow\}$ for the spin part of the spin orbitals.
All other spin blocks either vanish if the net spin of the upper indices and lower indices differs, e.g. 
\begin{align}
D^{i_1 \uparrow i_2 \uparrow}_{j_1 \uparrow j_2 \downarrow}=0,
\end{align}
or can be reconstructed using the antisymmetry of the 2-RDM and the spin-flip symmetry $\left(\uparrow \right) \leftrightarrow \left(\downarrow \right)$, e.g.,
\begin{align}
&D^{i_2 \downarrow i_1 \uparrow}_{j_1 \uparrow j_2 \downarrow}=-D^{i_1 \uparrow i_2 \downarrow}_{j_1 \uparrow j_2 \downarrow} 
&D^{i_1 \downarrow i_2 \downarrow}_{j_1 \downarrow j_2 \downarrow}=D^{i_1 \uparrow i_2 \uparrow}_{j_1 \uparrow j_2 \uparrow}.
\end{align}
Further symmetries based on Eq.~\ref{eq:spin_Sz} pose constraints on the contractions of the 2-RDM spin blocks. The vanishing norm of the vector $S_z\vert \Psi \rangle$ gives
\begin{align}\label{eq:weak_Sz_cond}
0&=\langle \Psi \vert S_z S_z\vert \Psi \rangle \nonumber\\
&=\frac{1}{2}\sum_{i,m}\left( D^{i \uparrow m \uparrow}_{i \uparrow m \uparrow} - D^{i \uparrow m \downarrow}_{i \uparrow m \downarrow}\right) + \frac{1}{2}\sum_i D^{i\uparrow}_{i\uparrow}\nonumber\\
&=\frac{N^2}{4}-\sum_{i,m} D^{i\uparrow m\downarrow}_{i\uparrow m\downarrow},
\end{align}
where we have used the interrelation between the 2-RDM and the 1-RDM (Eq.~\ref{eq:interrelation})
\begin{align}
\sum_m\left( D^{i\uparrow m\uparrow}_{j\uparrow m\uparrow} 
+ D^{i\uparrow m\downarrow}_{j\uparrow m\downarrow}\right)=(N-1)D^{i\uparrow}_{j\uparrow},
\end{align}
and
\begin{align}
\sum_i D^{i\uparrow}_{i\uparrow} = \frac{N}{2}.
\end{align}
Similarly, $S_z\vert \Psi \rangle=0$ implies
\begin{align}\label{eq:strong_Sz_cond}
0&=\langle \Psi \vert \hat{a}^\dagger_{i\uparrow}\hat{a}_{j\uparrow} S_z\vert \Psi \rangle \nonumber \\
&=\frac{N}{2}D^{i\uparrow}_{j\uparrow}-\sum_{m} D^{i\uparrow m\downarrow}_{j\uparrow m\downarrow}.
\end{align}
We note that Eq.~\ref{eq:strong_Sz_cond} reduces to Eq.~\ref{eq:weak_Sz_cond} by tracing out the non-contracted indices. While for the $N$-particle state $\Psi(t)$ the conditions $\langle\Psi|\hat S_z\hat S_z\vert \Psi \rangle=0$ and $\hat{S}_z\vert \Psi \rangle=0$ are equivalent, this is not the case for the (in general, non $N$-representable) 2-RDM within a truncated BBGKY hierarchy. For the latter, Eq.~\ref{eq:strong_Sz_cond} imposes additional constraints not implied by Eq.~\ref{eq:weak_Sz_cond}. \\
Further spin symmetries of the 2-RDM can be derived from $\hat{S}_+\vert \Psi \rangle=0$. The vanishing norm of the vector $\hat{S}_+\vert \Psi \rangle$ implies
\begin{align}\label{eq:weak_S2_cond}
0&=\langle \Psi \vert S_- S_+ \vert \Psi \rangle =\sum_{i}D^{i \uparrow}_{i \uparrow}-\sum_{i,m}D^{m \uparrow i \downarrow}_{i \uparrow m \downarrow}\nonumber\\
&=\frac{N}{2}-\sum_{i,m}D^{m \uparrow i \downarrow}_{i \uparrow m \downarrow},
\end{align}
and the stronger condition
\begin{align}\label{eq:strong_S2_cond}
0&=\langle \Psi \vert \hat{a}^\dagger_{i\downarrow}\hat{a}_{j\uparrow} S_+ \vert \Psi \rangle =D^{i \uparrow}_{j \uparrow}-\sum_{m}D^{m \uparrow i \downarrow}_{j \uparrow m \downarrow}.
\end{align}
We, furthermore, derive an interrelation between blocks of the 2-RDM. Projecting the vector $\hat{S}_+\vert \Psi \rangle$ onto two-particle-two-hole excitations we find 
\begin{align}
	0=&\langle \Psi \vert 
	\hat{a}^\dagger_{i_1\uparrow}
	\hat{a}^\dagger_{i_2\downarrow}
	\hat{a}_{j_2\uparrow} 
	\hat{a}_{j_1\uparrow}  
	S_+\vert \Psi \rangle \nonumber\\
	=&\sum_k \langle \Psi \vert 
	\hat{a}^\dagger_{i_1\uparrow}
	\hat{a}^\dagger_{i_2\downarrow}
	\hat{a}_{j_2\uparrow} 
	\hat{a}^\dagger_{k\uparrow}
	\hat{a}_{k\downarrow}
	\hat{a}_{j_1\uparrow} \vert \Psi \rangle - D^{i_1 \uparrow i_2 \downarrow}_{j_2 \uparrow j_1 \downarrow} \nonumber \\
	=&\sum_k\langle \Psi \vert 
	\hat{a}^\dagger_{i_1\uparrow}
	\hat{a}^\dagger_{i_2\downarrow}
	\hat{a}^\dagger_{k\uparrow}
	\hat{a}_{k\downarrow}
	\hat{a}_{j_2\uparrow} 
	\hat{a}_{j_1\uparrow}
	\vert \Psi \rangle - D^{i_1 \uparrow i_2 \downarrow}_{j_2 \uparrow j_1 \downarrow} +
	 D^{i_1 \uparrow i_2 \downarrow}_{j_1 \uparrow j_2 \downarrow}
	\nonumber\\
	=&\langle \Psi \vert 
	\hat{S}_+
	\hat{a}^\dagger_{i_1\uparrow}
	\hat{a}^\dagger_{i_2\downarrow}
	\hat{a}_{j_2\uparrow} 
	\hat{a}_{j_1\uparrow}
	\vert \Psi \rangle  \nonumber \\
	&- D^{i_1 \uparrow i_2 \uparrow}_{ j_1 \uparrow j_2 \uparrow} + D^{i_1 \uparrow i_2 \downarrow}_{j_1 \uparrow j_2 \downarrow}
		 - D^{i_1 \uparrow i_2 \downarrow}_{j_2 \uparrow j_1 \downarrow}\nonumber\\
	=&-D^{i_1 \uparrow i_2 \uparrow}_{j_1 \uparrow j_2 \uparrow}+D^{i_1 \uparrow i_2 \downarrow}_{j_1 \uparrow j_2 \downarrow}-D^{i_1 \uparrow i_2 \downarrow}_{j_2 \uparrow j_1 \downarrow}.\label{eq:singlet_cond}
\end{align}
Consequently, we arrive at the important interrelation 
\begin{align}\label{eq:antisym_Dab}
D^{i_1 \uparrow i_2 \uparrow}_{j_1 \uparrow j_2 \uparrow} = D^{i_1 \uparrow i_2 \downarrow}_{j_1 \uparrow j_2 \downarrow} - D^{i_1 \uparrow i_2 \downarrow}_{j_2 \uparrow j_1 \downarrow},
\end{align}
which has been derived previously employing the Wigner-Eckhard theorem \cite{WeeMiz61}. Equation \ref{eq:antisym_Dab} has a simple interpretation: Since $ D^{i_1 \uparrow i_2 \downarrow}_{j_1 \uparrow j_2 \downarrow} - D^{i_1 \uparrow i_2 \downarrow}_{j_2 \uparrow j_1 \downarrow}$ is antisymmetric with respect to the spatial indices it belongs to the spin triplet state of a pair in spin state $|S M_s\rangle = |10\rangle$. In terms of the spatial orbitals it has exactly the same eigenvectors and eigenvalues (except of a factor two) as $D^{i_1 \uparrow i_2 \uparrow }_{j_1 \uparrow j_2 \uparrow}$ which belongs to the spin state $|S M_s\rangle = |11\rangle$. Equation \ref{eq:antisym_Dab} is, therefore, also important for the numerical efficiency. Since the 2-RDM can be reconstructed completely from the $(\uparrow \downarrow)$-block, it is sufficient to propagate only the $(\uparrow \downarrow)$-block according to (see Eq.~\ref{eq:2p_eom_orbital})
\begin{align}\label{eq:eom_approxspin}
i \partial_t D^{i_1 \uparrow i_2 \downarrow}_{j_1 \uparrow j_2 \downarrow}=&\sum_{k_1,k_2}\big( H^{k_1  k_2 }_{j_1 j_2 }D^{i_1 \uparrow i_2 \downarrow}_{k_1 \uparrow k_2 \downarrow}-D^{k_1 \uparrow k_2 \downarrow}_{j_1 \uparrow j_2\downarrow}H^{i_1 i_2}_{k_1 k_2}\big)\nonumber\\
+&C^{i_1 \uparrow i_2 \downarrow}_{j_1 \uparrow j_2 \downarrow},
\end{align}
instead of the entire 2-RDM. $H^{k_1 k_2 }_{j_1 j_2 }$ are the matrix elements of the Hamiltonian (Eq.~\ref{eq:2p_ham}) in the spatial orbitals.
This significantly reduces the numerical effort because the $(\uparrow \downarrow)$-block of the collision operator can be written solely in terms of the $(\uparrow \uparrow \downarrow)$-block of the 3-RDM
\begin{align}
&C^{i_1 \uparrow i_2 \downarrow}_{j_1 \uparrow j_2 \downarrow}=
I^{i_1 \uparrow i_2 \downarrow}_{j_1 \uparrow j_2 \downarrow}
+I^{i_2 \uparrow i_1 \downarrow}_{j_2 \uparrow j_1 \downarrow}
-(I^{j_1 \uparrow j_2 \downarrow}_{i_1 \uparrow i_2 \downarrow}
+I^{j_2 \uparrow j_1 \downarrow}_{i_2 \uparrow i_1 \downarrow})^*,
\label{eq:def_Cop_spin}
\end{align}
and
\begin{align}
I^{i_1 \uparrow i_2 \downarrow}_{j_1 \uparrow j_2 \downarrow}=\sum_{k_1,k_2,k_3}W_{j_1 k_1}^{k_2 k_3}\big( D_{k_2\uparrow k_3\uparrow j_2\downarrow}^{i_1\uparrow k_1\uparrow i_2\downarrow} +D_{j_2\uparrow k_3\uparrow k_2\downarrow}^{i_2\uparrow k_1\uparrow i_1\downarrow}   \big),
\label{eq:eval_I_spin}
\end{align}
where we have used the spin flip symmetry between $(\uparrow)$ and $(\downarrow)$. Propagating only the $(\uparrow \downarrow)$-block, the evaluation of the collision operator scales like $(r)^7$ instead of $(2r)^7$ with the number of spatial orbitals $r$.\\
Since the equation of motion for the 2-RDM involves the 3-RDM we inquire now into constraints that spin conservation imposes on the 3-RDM. Starting with Eq.~\ref{eq:antisym_Dab} and taking the time-derivative, we find that the 3-RDM must fulfill
\begin{align}\label{eq:3rdm_spin_cond}
	0=D^{i_1 \uparrow i_2 \uparrow i_3 \uparrow}_{j_1 \uparrow  j_2 \uparrow j_3 \uparrow}
	+D^{i_1 \uparrow i_2 \uparrow i_3 \downarrow}_{j_1 \uparrow  j_2 \uparrow j_3 \downarrow}
	-D^{i_1 \uparrow i_3 \uparrow i_2 \downarrow}_{j_3 \uparrow  j_2 \uparrow j_1 \downarrow}\nonumber\\
	-D^{i_1 \uparrow i_3 \uparrow i_2 \downarrow}_{j_1 \uparrow  j_3 \uparrow j_2 \downarrow}
	-D^{i_2 \uparrow i_3 \uparrow i_1 \downarrow }_{j_3 \uparrow j_1 \uparrow j_2 \downarrow}
	-D^{i_2 \uparrow i_3 \uparrow i_1 \downarrow }_{j_2 \uparrow j_3 \uparrow j_1 \downarrow},
\end{align}
which is a consequence of 
\begin{align}
0=&\langle \Psi \vert 
	\hat{a}^\dagger_{i_1\uparrow}
	\hat{a}^\dagger_{i_2\uparrow}
	\hat{a}^\dagger_{i_3\downarrow}
	\hat{a}_{j_3\uparrow} 
	\hat{a}_{j_2\uparrow} 
	\hat{a}_{j_1\uparrow} S_+ \vert \Psi \rangle \nonumber \\
	+&
	\langle \Psi \vert 
		\hat{a}^\dagger_{i_1\downarrow}
		\hat{a}^\dagger_{i_2\downarrow}
		\hat{a}^\dagger_{i_3\uparrow}
		\hat{a}_{j_3\downarrow} 
		\hat{a}_{j_2\uparrow} 
		\hat{a}_{j_1\uparrow} S_+ \vert \Psi \rangle.
\end{align}
We focus in the following on the $(\uparrow \uparrow \downarrow)$-block of the 3-RDM. This block has four independent one-fold contractions. From the time derivative of Eq.~\ref{eq:strong_Sz_cond} we obtain conditions for two of these
\begin{align}
	&\sum_{m}D^{i_1 \uparrow m \uparrow i_2 \downarrow}_{j_1 \uparrow m \uparrow j_2 \downarrow}=\left(\frac{N}{2}-1\right)D^{i_1 \uparrow i_2 \downarrow}_{j_1 \uparrow j_2 \downarrow} \label{eq:3RDM_contr1}\\
	&\sum_{m}D^{i_1 \uparrow i_2 \uparrow m \downarrow}_{j_1 \uparrow j_2 \uparrow m \downarrow}=\frac{N}{2} D^{i_1 \uparrow i_2 \uparrow}_{j_1 \uparrow j_2 \uparrow},\label{eq:3RDM_contr2}
\end{align}
and by taking the time derivative of Eq.~\ref{eq:strong_S2_cond} the two remaining ones
 \begin{align}
 	&\sum_m D^{i_1 \uparrow m \uparrow i_2 \downarrow}_{j_1 \uparrow j_2 \uparrow m \downarrow}=D^{i_1 \uparrow i_2 \uparrow}_{j_1 \uparrow j_2 \uparrow} \label{eq:3RDM_contr3}\\
 	&\sum_m D^{i_1 \uparrow i_2 \uparrow m \downarrow}_{j_1 \uparrow m \uparrow j_2 \downarrow}=D^{i_1 \uparrow i_2 \uparrow}_{j_1 \uparrow j_2 \uparrow}.\label{eq:3RDM_contr4}
 \end{align}
In analogy to energy conservation, we find that conservation of spin requires that a properly reconstructed 3-RDM correctly contracts in all diagonal and off-diagonal partial traces to the 2-RDM. These are important constraints on the reconstruction functionals of the 3-RDM, $D^{\rm R}_{123}[D_{12}]$, unfortunately not fulfilled by reconstruction functionals previously discussed in the literature.
\section{\label{sec:con}Contraction-consistent reconstruction of the 3-RDM}
%
The approximate reconstruction of higher-order RDMs in terms of lower order RDMs has been successfully developed in the last few decades to remove the indeterminacy of the time-independent contracted Schr\"odinger equation which depends on both the 3-RDM and the 4-RDM (see, e.g., \cite{ColPerVal93,YasNak97,Mazz98,Maz00_complete}). In a pioneering work exploiting particle-hole duality \cite{ColPerVal93}, the following reconstruction functional for the 3-RDM referred to as the Valdemoro (V) reconstruction functional
\begin{align}\label{eq:val}
&D^{\mathrm V}_{123}[D_{12}]=9 D_{12} \wedge D_1 - 12 D_1^3
\end{align} 
has been derived, where the wedge product is defined as the antisymmetrized tensor product
\begin{align}\label{eq:wedge}
&D_{1\dots p} \wedge D_{1\dots q}=
\nonumber \\
&=\frac{1}{(p+q)!^2} \sum_{\pi,\tau}\text{sgn}(\pi) \text{sgn}(\tau)
D^{i_{\pi(1)}...i_{\pi(p)}}_{j_{\tau(1)}...j_{\tau(p)}}
D^{i_{\pi(p+1)}...i_{\pi(p+q)}}_{j_{\tau(p+1)}...j_{\tau(p+q)}},
\end{align}
and the sum runs over the permutations $\pi$ and $\tau$. The error in the reconstruction 
\begin{align}\label{eq:cumulant}
\Delta_{123}=D_{123}-D^{\mathrm V}_{123}[D_{12}]
\end{align}
is the three-particle cumulant $\Delta_{123}$ as has been pointed out in \cite{Maz98}. It describes the part of the 3-RDM that cannot be constructed from $D_{12}$ and $D_1$. Physically, neglecting the three-particle cumulant amounts to neglecting all processes that cannot be viewed as a sequence of independent two and one-particle excitations. In the following we will show that the assumption of a vanishing three-particle cumulant, i.e.~reconstruction via $D_{123}^{\mathrm V}$, leads to the violation of spin and energy conservation. It is thus essential to include parts of the cumulant in the reconstruction in order to preserve these conservation laws.\\
Remarkably, the Valdemoro reconstruction functional conserves the weaker conditions Eq.~\ref{eq:weak_Sz_cond} and Eq.~\ref{eq:weak_S2_cond} as well as Eq.~\ref{eq:antisym_Dab} but violates the stronger conditions Eq.~\ref{eq:strong_Sz_cond} and Eq.~\ref{eq:strong_S2_cond}, since $D_{123}^{\mathrm V\uparrow  \uparrow \downarrow}$ does not contract correctly into the two-particle subspace according to Eq.~\ref{eq:3RDM_contr1} - Eq.~\ref{eq:3RDM_contr4}. In other words the failure of the Valdemoro reconstruction $D_{123}^{\mathrm V \uparrow  \uparrow \downarrow}$ originates from the fact that the three-particle cumulant $\Delta_{123}^{\uparrow  \uparrow \downarrow}$ has non-vanishing contractions, e.g.
\begin{align}\label{eq:cont_cumul}
&\sum_{m}\Delta^{i_1 \uparrow i_2 \uparrow m \downarrow}_{j_1 \uparrow j_2 \uparrow m \downarrow}=\frac{N}{2}D^{i_1 \uparrow i_2 \uparrow}_{j_1 \uparrow j_2 \uparrow}-\sum_{m}[D^\mathrm{V}]^{i_1 \uparrow i_2 \uparrow m \downarrow}_{j_1 \uparrow j_2 \uparrow m \downarrow}\neq 0.
\end{align}
The information on the cumulant stored in the 2-RDM can be used to develop a new contraction-consistent reconstruction functional that satisfies (Eq.~\ref{eq:3RDM_contr1} - Eq.~\ref{eq:3RDM_contr4}) and, therefore, ensures spin and energy conservation. For this purpose we decompose the three-particle cumulant 
\begin{align}\label{eq:cumul}
\Delta_{123}^{\uparrow  \uparrow \downarrow}=\Delta_{123;\perp}^{\uparrow  \uparrow \downarrow}[D_{12}]+\Delta_{123;\text{K}}^{\uparrow  \uparrow \downarrow}
\end{align}
into the contraction-free component $\Delta_{123;\text{K}}^{\uparrow  \uparrow \downarrow}$ and the corresponding orthogonal component $\Delta_{123;\perp}^{\uparrow  \uparrow \downarrow}$ using the unitary decomposition for three particle matrices described in Appendix \ref{sec:uni}.
By definition, the contraction-free component vanishes upon all diagonal and off-diagonal contractions denoted by $L_3$
\begin{align}
L_3(\Delta_{123;\text{K}}^{\uparrow  \uparrow \downarrow})=0,
\end{align}
and is thus an element of the kernel of $L_3$. The key ingredient for the contraction-consistent reconstruction is the fact that the orthogonal component of the cumulant $\Delta_{123;\perp}^{\uparrow  \uparrow \downarrow}[D_{12}]$ is exactly given as a functional of the 2-RDM. Using $\Delta_{123;\perp}^{\uparrow  \uparrow \downarrow}$, we obtain the new contraction-consistent reconstruction
\begin{align}\label{eq:cc_recon}
D_{123}^{C \uparrow  \uparrow \downarrow}[D_{12}] = D_{123}^{\mathrm V \uparrow  \uparrow \downarrow}[D_{12}] + \Delta_{123;\perp}^{\uparrow  \uparrow \downarrow}[D_{12}],
\end{align}
which satisfies Eq.~\ref{eq:3RDM_contr1} - Eq.~\ref{eq:3RDM_contr4}. It differs from the exact 3-RDM only by the contraction-free component $\Delta_{123;\text{K}}^{\uparrow  \uparrow \downarrow}$. This is to be compared with the Valdemoro reconstruction functional (Eq.~\ref{eq:val}) that neglects the cumulant altogether. We note that despite this improvement, the contraction-consistent reconstruction functional is not sufficient to ensure $N$-representability of the 2-RDM during time evolution.

\section{\label{sec:N-r}$N$-representability and dynamical purification}
%
Each $N$-particle density matrix that is (i) hermitian, (ii) normalized, (iii) antisymmetric under particle permutation, and (iv) positive semidefinite describes a possible state of a $N$-particle system. The conditions for the $p$-RDM to describe a $p$-particle subsystem of the original $N$-particle system are much more complex. Subsidiary conditions have to be imposed to ensure that the RDM belongs to an actual wavefunction. This conditions are called $N$-representability conditions \cite{Col63}. The search for a complete set of conditions for the 2-RDM is an ongoing effort for over half a century \cite{GarRos69,Erd79,ZhaBraOve03,Col63}. A systematic classification of $N$-representability conditions has been developed \cite{Maz12} for ensemble representable RDMs, i.e., matrices that are derivable from a mixed quantum state. The actual form of a complete set of conditions for pure states remains undetermined. Moreover, numerical calculations allow to implement only few $N$-representability conditions. There exist several explicit necessary conditions for $N$-representability in the form of positivity conditions. The two most important positivity conditions for the 2-RDM are called the D and the Q-positivity condition \cite{Maz07}. They guarantee that the 2-RDM
\begin{align} \label{eq:D_cond}
D^{i_1i_2}_{j_1j_2}&=\langle \Psi \vert \hat{a}^\dagger_{i_1} \hat{a}^\dagger_{i_2} \hat{a}_{j_2}\hat{a}_{j_1} \vert \Psi \rangle
\end{align}
and the two-hole reduced density matrix (2-HRDM)
\begin{align}
Q^{i_1i_2}_{j_1j_2}&=\langle  \Psi \vert \hat{a}_{j_1}\hat{a}_{j_2} \hat{a}^\dagger_{i_2} \hat{a}^\dagger_{i_1} \vert \Psi \rangle \label{eq:Q_cond}
\end{align}
are positive semidefinite (i.e.~have non-negative eigenvalues). The 2-HRDM describes the pair distribution of holes rather than of particles. The positive semidefiniteness of these matrices represents independent conditions although the matrices are interconvertible by a rearrangement of the creation and annihilation operators
\begin{align}
Q_{12} &= 2 I^2-4 I \wedge D_1 + D_{12}.
\end{align}
These 2-positivity conditions imply that the occupation numbers of particle pairs or hole pairs in any two-particle state are always non-negative. A third 2-positivity condition, the G-condition, guarantees that the occupation of particle-hole pairs is non-negative \cite{Maz07}. For the calculations presented here, the G-condition turned out to be much less important than the D and Q-condition. In fact, the G-condition was found to be well conserved whenever the D and Q-condition were fulfilled. The 2-positivity conditions are conveniently implemented since they can be formulated solely in terms of the 2-RDM.
\begin{figure}
\centering
\includegraphics[width=1\linewidth]{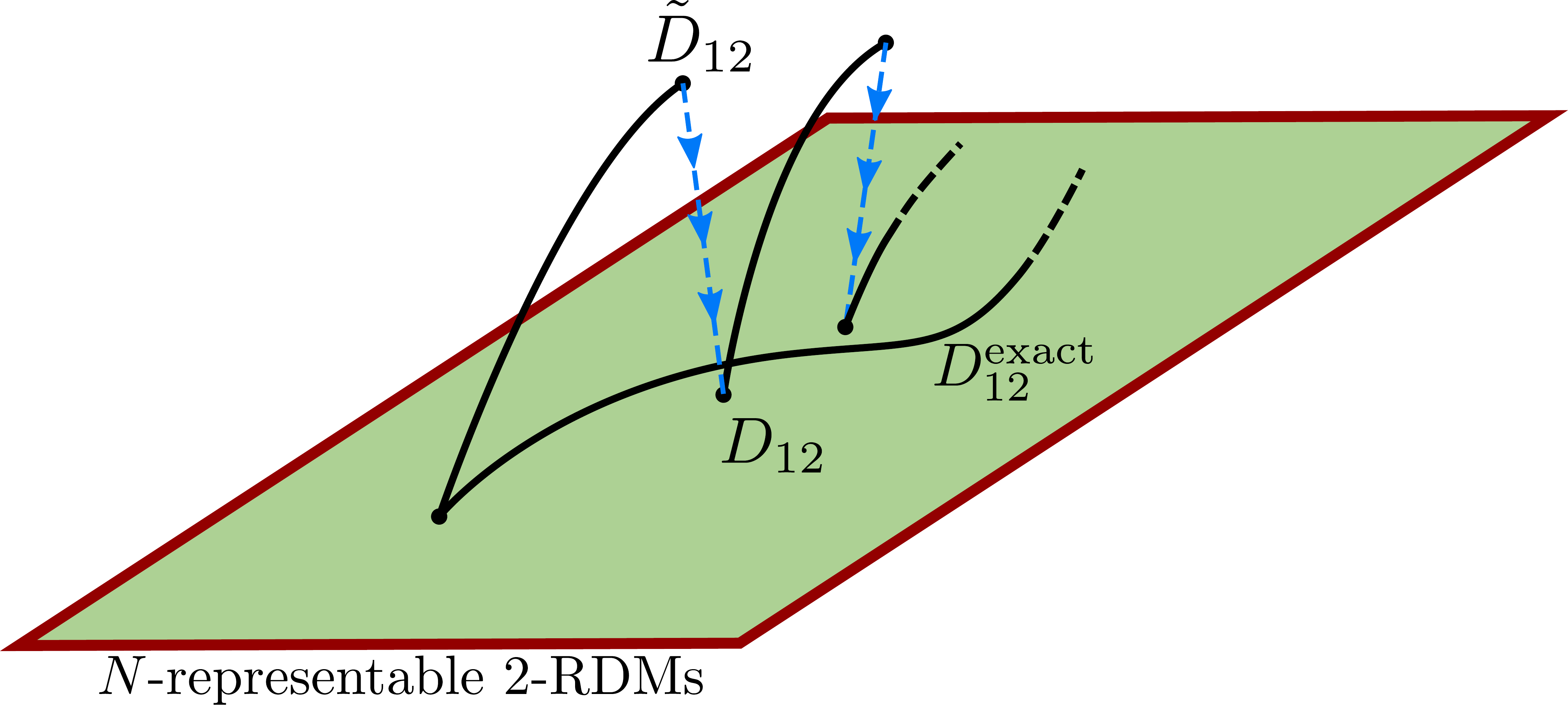}
\caption{(Color online) Dynamical purification applied after each time step to project the propagated $\tilde{D}_{12}(t+\Delta t)$ onto the set of 2-RDMs that satisfy the D and Q-condition, schematically.}
\label{fig:purify2}
\end{figure}
Even when the 2-RDM associated with the initial state satisfies the D and the Q-positivity condition the $N$-representability conditions may be violated during time evolution calculated according to Eq.~\ref{eq:eom_approx} due to the residual errors in the reconstruction functional. The 2-RDM $D_{12}(t)$ will, in general, depart from the subspace of $N$-representable 2-RDMs after propagation for the time step $\Delta t$. Therefore, projecting back the evolved $\tilde{D}_{12}(t+\Delta t)$ onto the subspace of $N$-representable 2-RDMs,
\begin{align}\label{eq:projection}
D_{12}(t+\Delta t)= \hat{P}_{12}\tilde{D}_{12}(t+\Delta t)\hat{P}_{12},
\end{align}
where the projector $\hat{P}_{12}$ enforces a set of preselected representability conditions is essential (Fig.~\ref{fig:purify2}). This process is referred to in the following as dynamical purification. Several types of purifications have been discussed in literature and are used primarily for the iterative solution of the second order contracted Schr\"odinger equation to find a self-consistent $N$-representable solution for the ground state of molecules \cite{Maz02Pur,AlcCasVal05}. \\
A purification scheme which accounts for the D and the Q-condition and employs the unitary decomposition (see Eq.~\ref{eq:decom_2RDM} in the Appendix) \cite{Maz02Pur} serves as the starting point of our purification process for the time-dependent 2-RDM. Briefly, we add to both the $\tilde{D}_{12}(t)$ and the $\tilde{Q}_{12}(t)$ a correction term 
\begin{align}
D_{12}(t)=\tilde{D}_{12}(t)+D_{12}^{\text{cor}}(t)\label{eq:purif_corr}\\
Q_{12}(t)=\tilde{Q}_{12}(t)+D_{12}^{\text{cor}}(t)
\end{align}
with
\begin{align}\label{eq:correct}
D_{12}^{\text{cor}}(t)=\sum_i \big(\alpha_i A^i_{12;\text{K}}+\beta_i B^i_{12;\text{K}}\big).
\end{align}
In Eq.~\ref{eq:correct} the $A^i_{12;\text{K}}$ and $B^i_{12;\text{K}}$ are the contraction free components (see Eq.~\ref{eq:decom_2RDM}) of the projections onto the geminals (i.e.~the eigenvectors) $A^i_{12}=\vert g_i \rangle \langle g_i \vert$ and $B^i_{12}=\vert g'_i \rangle \langle g'_i \vert$ with negative eigenvalues of $\tilde{D}_{12}$ or $\tilde{Q}_{12}$, respectively. In order to preserve the D and Q-positivity condition the negative eigenvalues are reduced by solving the system of linear equations for the coefficients $\alpha_i$ and $\beta_i$
\begin{align}\label{eq:pur_2RDM1}
\text{Tr}_{12}(A^i_{12} D_{12})&=0 \\\label{eq:pur_2RDM2}
\text{Tr}_{12}(B^i_{12} Q_{12})&=0.
\end{align}
Correcting the 2-RDM via Eq.~\ref{eq:purif_corr} creates a new $D_{12}$ with preserved 1-RDM, and whose negative eigenvalues are smaller than those of $\tilde{D}_{12}$. Repeating this process iteratively yields the purified $D_{12}$(t). We note that this iterative procedure converges only if the underlying 1-RDM is $N$-representable, i.e.~has eigenvalues between 0 and 1. We find that the time-dependent 1-RDM remains $N$-representable during the evolution when the D and Q-condition on $D_{12}$ and $Q_{12}$ are enforced.\\
The purification process outlined above requires modification when spin symmetries are to be conserved simultaneously. We first note that it is sufficient to only purify the $(\uparrow\downarrow)$-block because in the singlet spin state this block contains all the information of the full 2-RDM and has the same eigenvalues (except of a factor two) as the full 2-RDM. The D and Q-condition are then equivalent to the positivity of the $(\uparrow \downarrow)$-block of the 2-RDM and the 2-HRDM. We separate the $(\uparrow \downarrow)$-block further into the symmetric and the antisymmetric part with respect to the spatial orbital indices
\begin{align}
D^{\uparrow \downarrow}_{12}&=\hat{\mathcal{A}}[D^{\uparrow \downarrow}_{12}]+\hat{\mathcal{S}}[D^{\uparrow \downarrow}_{12}] \label{eq:pur_spin_1}\\
Q^{\uparrow \downarrow}_{12}&=\hat{\mathcal{A}}[Q^{\uparrow \downarrow}_{12}]+\hat{\mathcal{S}}[Q^{\uparrow \downarrow}_{12}], \label{eq:pur_spin_2}
\end{align}
where $\hat{\mathcal{A}}$ is the antisymmetrization operator and $\hat{\mathcal{S}}$ is the symmetrization operator. While for the antisymmetric part we can directly apply the purification described above, the purification of the symmetric part employs the unitary decomposition for symmetric matrices (see Eq.~\ref{eq:ansatz_d12_sym} in the Appendix). 
This purification does not alter the one-particle traces of the $(\uparrow \downarrow)$-block such that the conditions Eq.~\ref{eq:weak_Sz_cond}, Eq.~\ref{eq:strong_Sz_cond}, Eq.~\ref{eq:weak_S2_cond}, and Eq.~\ref{eq:strong_S2_cond} remain conserved. The convergence is strongly dependent on the positive semidefiniteness of the one-particle traces of the symmetric and antisymmetric components:
\begin{align}\label{eq:conv_cond}
&\mathrm{Tr}_{2}\hat{\mathcal{A}}[D^{\uparrow \downarrow}_{12}] \ge 0
&\mathrm{Tr}_{2}\hat{\mathcal{A}}[Q^{\uparrow \downarrow}_{12}] \ge 0 \nonumber\\
&\mathrm{Tr}_{2}\hat{\mathcal{S}}[D^{\uparrow \downarrow}_{12}] \ge 0
&\mathrm{Tr}_{2}\hat{\mathcal{S}}[Q^{\uparrow \downarrow}_{12}] \ge 0.
\end{align}
If conditions Eq.~\ref{eq:strong_Sz_cond} and Eq.~\ref{eq:strong_S2_cond} are met these matrices are proportional to the 1-RDM and, therefore, positive semidefinite. In the propagation with $D_{123}^{\mathrm V}$ these conditions are violated causing convergence problems of the spin adapted purification. For test calculations with $D_{123}^{\mathrm V}$ we, therefore, use the purification of the whole 2-RDM (Eq.~\ref{eq:purif_corr}) rather than the purification of the $(\uparrow\downarrow)$-block. In the following we show that the dynamical purification process is key to achieve a stable propagation of the 2-RDM.

\section{\label{sec:orb}Benchmark: LiH in a few-cycle laser field}
%
In this section we present a first application of our TD-2RDM method to a four-electron model system, the one-dimensional LiH molecule in an ultrashort few-cycle laser field.
One-dimensional atoms and molecules serve as a numerically efficient testing ground for full three-dimensional (3D) calculations and have been used in the past to study various atomic properties such as the double ionization of He \cite{DahLee01} and H$_2$ \cite{HenZuoBan96} and the response of LiH in strong laser fields \cite{BalBauBon10,TakKen13}. We have chosen this system since it displays already a complex and rich multi-electron dynamics, including multiple ionization, while it still can be numerically exactly solved employing the MCTDHF method allowing to accurately benchmark the TD-2RDM method. For the numerical implementation, we solve the orbital equations of motion (Eq.~\ref{eq:dphi_MCTDH}) on an equidistant grid with $2000$ points and grid spacing $\Delta z =0.1$ for the laser intensity $I=10^{14}\rm{W/cm^2}$ and $3000$ points and grid spacing $\Delta z =0.4$ for the higher laser intensity $I=8\times 10^{14}\rm{W/cm^2}$. The second derivative of the kinetic energy operator is evaluated within the eighth-order finite difference representation. An absorbing boundary is implemented by the mask function of $\cos^{\frac{1}{4}}$ shape. We employ the Runge-Kutta propagator of fourth order to propagation in real and imaginary time, the latter for the determination of the ground state.

\subsection{The LiH ground state}\label{sec:gro}

\begin{figure}
\centering
\includegraphics[width=\linewidth]{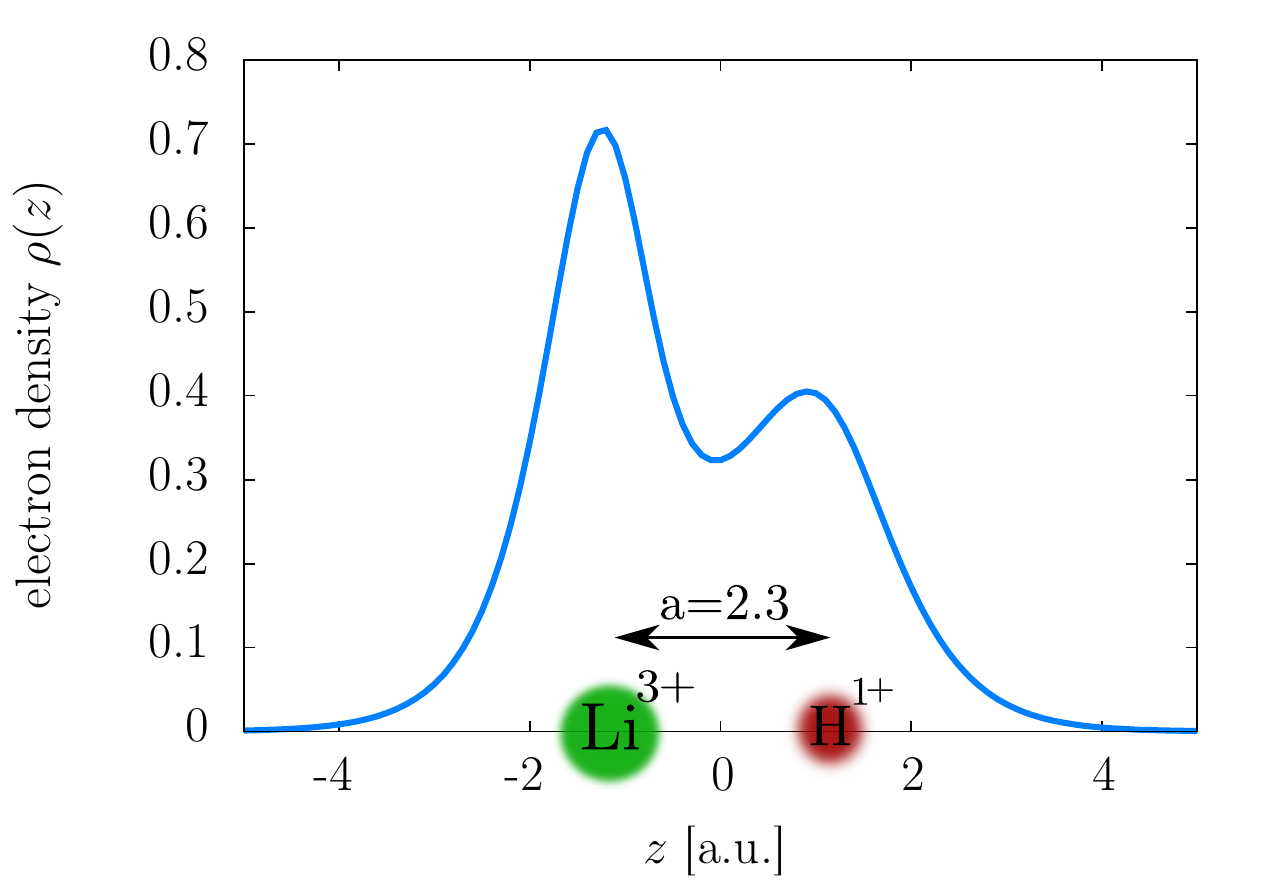}
\caption{(Color online) Electron density of the 1D LiH molecule. The equilibrium bond length $a=2.3$ between the Li nucleus and the proton is depicted. The electron cloud is predominantly located near the Li core.}
\label{fig:density}
\end{figure}

\begin{figure}
\centering
\includegraphics[width=\linewidth]{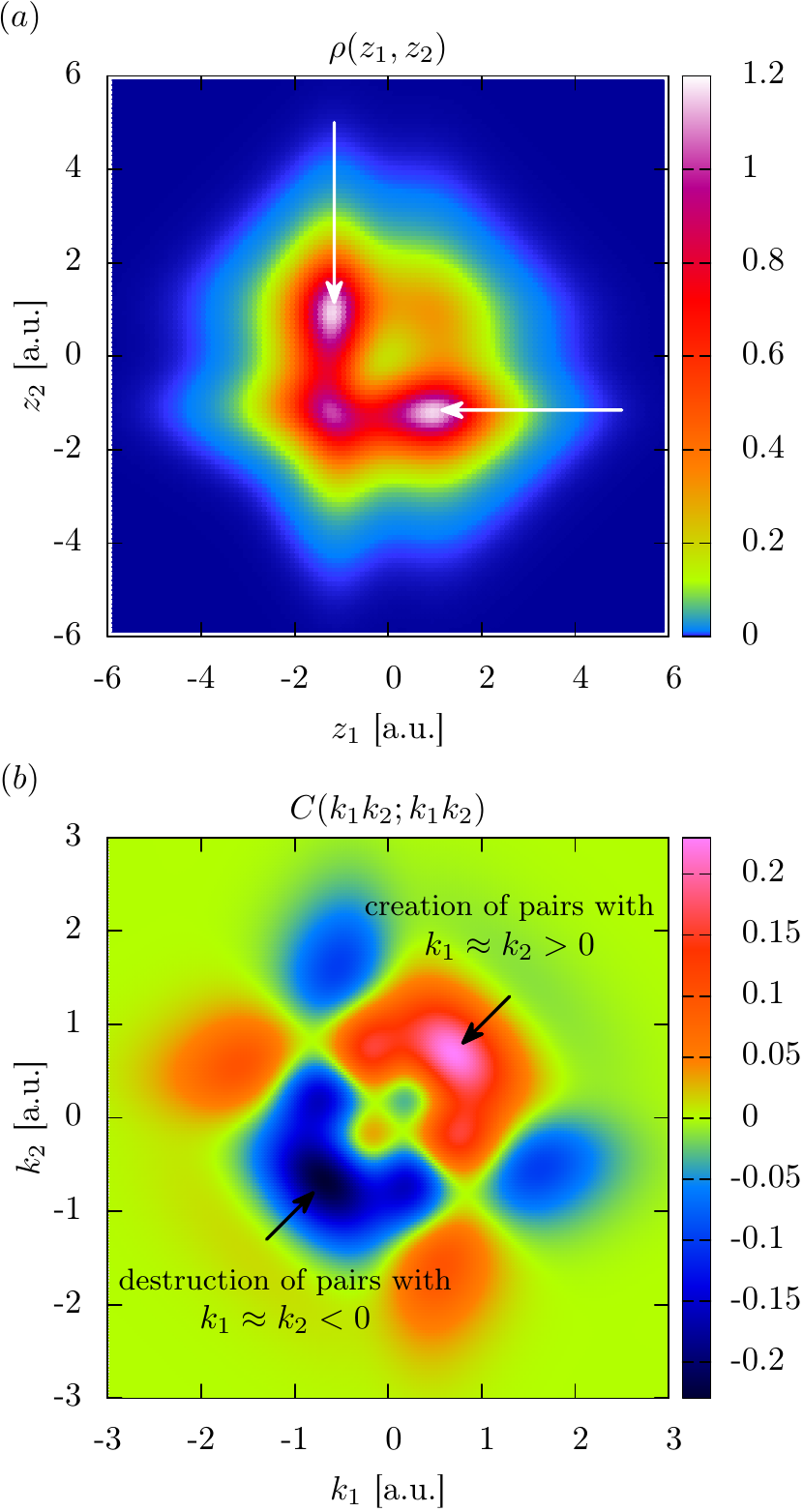}
\caption{(Color online) (a) The pair-density distribution $\rho(z_1,z_2)$ in coordinate space for the ground state of the LiH molecule. The density distribution shows distinct peaks for inter-atomic pairs with one electron close to the Li core while the other one is close to the proton (marked by arrows). (b) The imaginary diagonal elements of the exact collision operator $C_{12}$. The positive contribution for pairs with positive total momentum shows that the particle interaction creates pairs moving collectively toward the proton while pairs with negative momentum moving towards the Li core are destroyed.}
\label{fig:2RDM_ground}
\end{figure}

The electronic Hamiltonian (Eq.~\ref{eq:2p_ham}) of the one-dimensional model consists of the Li nucleus (charge $Z_{\rm Li}=3$) and the proton (charge $Z_{\rm H}=1$) at fixed positions $R_{\rm Li}$ and $R_{\rm H}$, and 4 electrons in the laser field $F(t)$ included within the dipole approximation in length gauge:
\begin{align}\label{eq:hamiltonian}
&H= \sum_{i=1}^4 h_i+\sum_{i<j}^4 W_{ij},
\end{align}
with
\begin{align}\label{eq:hi}
	h_i =-\frac{1}{2} \frac{\partial^2}{\partial z_i^2}+V_i+z_iF(t),
\end{align}
and the electron-electron interaction in 1D with softening parameter $d$
\begin{align}\label{eq:w12}
	W_{12} = \frac{1}{\sqrt{(z_1-z_2)^2 +d}}.
\end{align}
The one-electron molecular potential in Eq.~\ref{eq:hi} is given in 1D by 
\begin{align}\label{eq:vi}
V_i=- &\frac{Z_{\rm Li}}{\sqrt{(z_i-R_{\rm Li})^2+c}}-\frac{Z_{\rm H}}{\sqrt{(z_i-R_{\rm H})^2+c}}.
\end{align}
The softening parameters are chosen $c=0.5$ and $d=1$ with an equilibrium distance $a=\vert R_{\rm Li}- R_{\rm H}\vert=2.3$ ($R_{\rm Li}=-1.15$ and $R_{\rm H}=1.15$) \cite{TakKen13}. The ground state calculation which serves as the initial state for the TD-2RDM method is calculated using imaginary time propagation within MCTDHF. The electron density  expressed in terms of diagonal elements of the 1-RDM,
\begin{align}\label{eq:density}
\rho(z,t)&=D(z\uparrow;z\uparrow;t)+D_1(z\downarrow;z\downarrow;t),
\end{align}
of the ground state displays a distinct maximum near the Li atom (see Fig.~\ref{fig:density}) which originates from the deeply bound doubly occupied core orbital of Li. The outer two electrons occupy the valence orbital which is responsible for the chemical bond. Note that this single configuration picture serves only for qualitative illustration while the numerical simulation includes configuration interaction. From the electron density one obtains one-electron observables such as the dipole moment
\begin{align}\label{eq:dipole}
d(t)=d_0 -\int z \rho(z,t) \text{d}z ,
\end{align}
which consists of the static nuclear dipole moment $d_0=-2.3$, and the time-dependent electronic contribution. 
Two-particle properties beyond those derivable from the electron density can be calculated via the pair density $\rho(z_1,z_2,t)$ derived from the 2-RDM,
\begin{align}\label{eq:pair-den}
\rho(z_1,z_2,t)&=D(z_1\uparrow z_2\uparrow;z_1\uparrow z_2\uparrow;t)\nonumber\\
&+D(z_1\uparrow z_2\downarrow;z_1\uparrow z_2\downarrow;t) \nonumber\\
&+D(z_1\downarrow z_2\uparrow;z_1\downarrow z_2\uparrow;t) \nonumber\\
&+D(z_1\downarrow z_2\downarrow;z_1\downarrow z_2\downarrow;t).
\end{align}
The pair density $\rho(z_1,z_2,t)$ contains information beyond that of the electron density since it is influenced by two-particle correlations [see Fig.~\ref{fig:2RDM_ground} (a)]. In the LiH molecule the electron pairs are predominantly distributed such that one electron is located near the Li core and the other near the H core. This configuration minimizes the Coulomb repulsion. With the pair density $\rho(z_1,z_2,t)$ the exact interaction energy
\begin{align}\label{eq:E_int}
&E_{\text{int}}(t)= \int \frac{\rho(z_1, z_2, t)}{\sqrt{(z_1-z_2)^2+d}} \; \text{d}z_1 \text{d}z_2
\end{align}
including the full correlation energy can be calculated.
\\
A crucial quantity of the 2-RDM propagation is the collision operator $C_{12}$ (Eq.~\ref{eq:C12}). The collision operator is an antihermitian operator that describes the scattering between pairs under the influence of surrounding particles. More precisely, the imaginary part of the diagonal element $\langle \phi_{12} \vert C_{12} \vert \phi_{12} \rangle$ determines the number of pairs per unit time that enter minus those that leave the two-particle state $\vert \phi_{12} \rangle$ due to the Coulomb interaction with the $(N-2)$ electron environment. The diagonal elements in momentum space $C(k_1,k_2,k_1,k_2)$ can be directly interpreted as a Boltzmann-like collision integral. Since the wavefunction of a non-degenerate ground state is real the collision operator in momentum space is antisymmetric under point reflection at the origin [see Fig.~\ref{fig:2RDM_ground} (b)]
\begin{align}
C(k_1k_2;k'_1k'_2)&=C^*(-k_1,-k_2;-k'_1,-k'_2) \nonumber \\
&=-C(-k'_1,-k'_2;-k_1,-k_2).
\end{align}
If the system features in addition reflection symmetry in real space (e.g.~the beryllium atom), the collision operator must be also invariant under the transformation $(k_1,k_2) \rightarrow (-k_1,-k_2)$. This implies that the diagonal elements of the collision operator in momentum representation $C(k_1k_2;k_1k_2)$ must vanish for the ground state of such systems. We note that in coordinate space the diagonal elements of the collision operator always vanish, i.e.~$C(x_1x_2;x_1x_2)=0$. The fact that $C(k_1k_2;k_1k_2)$ does not vanish for LiH is a direct consequence of the broken parity symmetry of the LiH molecule. The ground state properties of the collision operator in momentum space can be understood intuitively by considering the equation of motion for the 2-RDM in momentum representation:
\begin{align}
i \partial_t D(k_1k_2;k'_1k'_2;t)=[H_{12},D_{12}](k_1k_2;k'_1k'_2;t)\\
+C(k_1k_2;k'_1k'_2;t).
\end{align}
The stationarity of the 2-RDM in the ground state imposes the condition
\begin{align}\label{eq:antiH_contr}
[H_{12},D_{12}](k_1k_2;k'_1k'_2;t)+C(k_1k_2;k'_1k'_2;t)=0.
\end{align}
Equation \ref{eq:antiH_contr} is equivalent to the antihermitian contracted stationary Schr\"odinger equation for the 2-RDM used to calculate the ground state of molecules (see e.g.~\cite{Maz06_antiher}). The momentum-space representation offers a straightforward interpretation in terms of a balance equation. The commutator $[H_{12},D_{12}]$ describes the change in the momentum distribution of electron pairs in the Coulomb field of the nuclei without the influence of the residual particles. The electron pairs are attracted collectively towards the Li core. This motion is compensated for by $C(k_1k_2;k_1k_2)$ which accounts for the collisions of the pairs with the surrounding particles which most prominently occurs at the enhanced electron density near the Li core driving the pairs toward the H core. This effect is visible as a maximum for positive total momenta and minimum for negative total momenta [see Fig.~\ref{fig:2RDM_ground} (b)]. 
In the stationary state these two competing processes are in equilibrium, i.e., for every pair that leaves the momentum configuration $(k_1,k_2)$ due to the interaction with the environment the core potential creates another pair of this kind.  

\subsection{The 2-RDM for LiH in an intense laser field} \label{sec:LiH}
%
\begin{figure}
\centering
\includegraphics[width=1\linewidth]{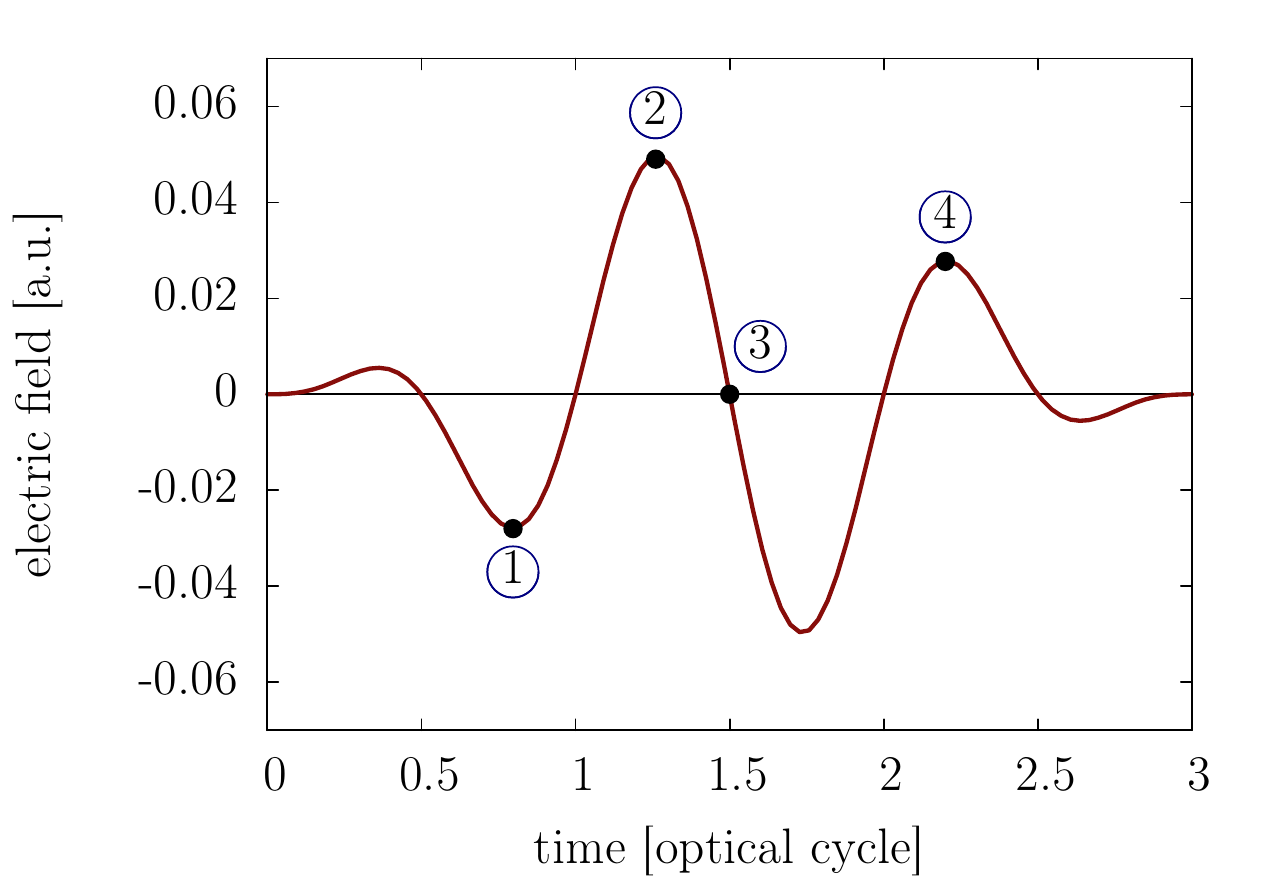}
\caption{(Color online) The laser pulse (Eq.~\ref{eq:laser_pulse}) with $I=10^{14}\rm{W/cm^2}$ ($F_0=0.053$), $\lambda=750$ nm, $N_c=3$. Distinct points in time are marked by numbers for later reference.}
\label{fig:pulse_form}
\end{figure}
For the  ultrashort few-cycle laser pulse we choose
\begin{align}\label{eq:laser_pulse}
F(t)=F_0 \sin(\omega t) \sin^2\left( \frac{\omega}{2N_c} t\right)  \quad 0 \le t \le N_c \frac{2\pi}{\omega},
\end{align}
where $F_0$ is the amplitude of the electric field, $\omega$ is the mean angular frequency, and $N_c$ is the number of cycles. We use from now on the scaled time $\tau=t  \frac{2\pi}{\omega}$ with $0 \le \tau \le N_c$, counting the number of cycles that have passed. We investigate two different laser intensities: $I=10^{14}\rm{W/cm^2}$ ($F_0=0.053$), for which the response of the dipole moment is close to linear, and $I=8\times10^{14}\rm{W/cm^2}$ ($F_0=0.151$) where a strongly nonlinear response is expected including substantial ionization. The Keldysh parameter 
\begin{align}
\gamma=\omega  \frac{\sqrt{2I_{\rm p}}}{F_0},
\end{align}
with the first ionization potential $I_{\rm p}=0.675$ (see \cite{TakKen13}) is $\gamma=1.32$ and $\gamma=0.467$, respectively. For all numerical results presented in this section the 2-RDM as well as the MCTDHF wavefunction $\Psi^{\rm MCTDHF}(t)$, with which we compare, are expanded in terms of 10 time-dependent spin orbitals (see Eq.~\ref{eq:exp_D_2}). The latter has been shown to be sufficient to reach convergence for the observables deduced from the MCTDHF wavefunction which we refer to as ``exact " results in the following \cite{TakKen13}. \\
We illustrate and assess now the accuracy of the present time-dependent 2-RDM theory by involving successively different levels of approximation to the collision operator $C_{12}$ and the equation of motion for the 2-RDM whose exact form is given by Eq.~\ref{eq:eom_d12} while the approximate form involving the reconstruction functional is given by Eq.~\ref{eq:eom_approx}.
As a figure of merit for the comparison with the exact calculation as well as other approximate methods we use the time-dependent dipole moment $d(t)$, a one-particle observable for which an explicit functional in terms of the time-dependent density $\rho(z,t)$ exists (Eq.~\ref{eq:dipole}) and which can thus be determined from effective mean-field theories such as TDDFT or TDHF without invoking any only approximately known read-out functional.  For other observables, the construction of the approximate read-out functional from the propagated density $\rho(z,t)$ remains a challenge \cite{LapLee98,RohSimBur06,BaxKir13}. Point of departure is the parallel propagation of the 2-RDM according to Eq.~\ref{eq:eom_approx} and the MCTDHF wavefunction $\Psi^{\rm MCTDHF}(t)$ from which the time-dependent $p$-RDMs for each time step can be exactly determined. Because of the expansion of the $p$-RDMs in time-dependent spin orbitals, the coupling of the evolution of the orbitals (Eq.~\ref{eq:dphi_MCTDH}) to that of the 2-RDM must also be accounted for when the hierarchy of approximations to the collision integral is explored.
\subsubsection{Test of the reconstruction functional}\label{sec:fir}
\begin{figure}[t]
\centering
\includegraphics[width=\linewidth]{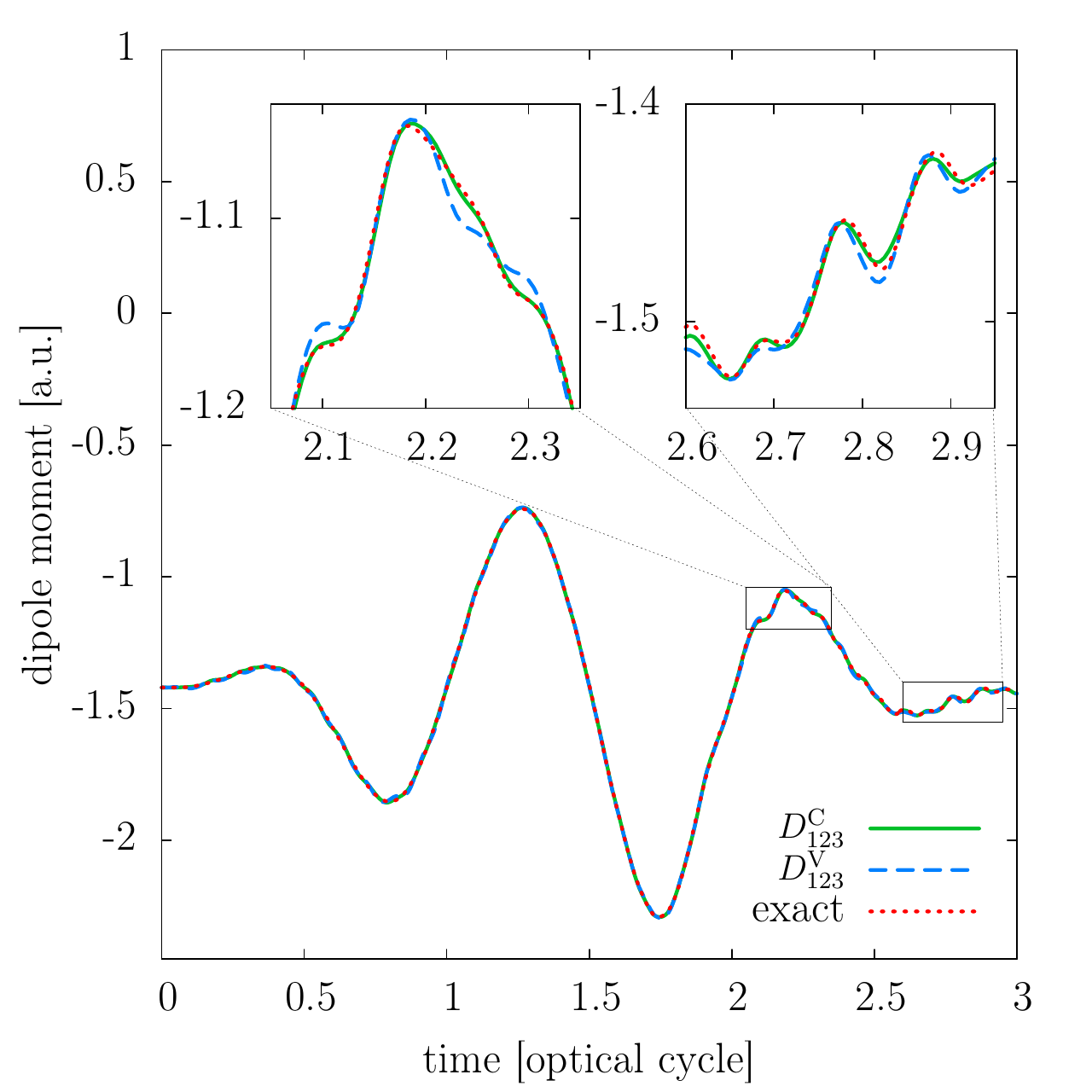}
\caption{(Color online) Dipole moment of LiH subject to the laser pulse as depicted in Fig.~\ref{fig:pulse_form} with $I=10^{14}\rm{W/cm^2}$ using the reconstruction functionals $D_{123}^{\mathrm V}[D^{\rm exact}_{12}]$ and $D_{123}^{\mathrm C}[D^{\rm exact}_{12}]$ with the exact 2-RDM $D^{\rm exact}_{12}$ as input at each time step obtained from a concurrent MCTDHF calculation. The high-frequency oscillations near $\tau=3$ originate from superpositions between the ground state and excited states (see right inset). Both $D_{123}^{\mathrm V}$ and $D_{123}^{\mathrm C}$ can markedly reproduce the MCTDHF result. A close up shows that $D_{123}^{\mathrm C}$ performs better than $D_{123}^{\mathrm V}$.}
\label{fig:first_level}
\end{figure}

The exact collision operator $C_{12}$ depends on the exact 3-RDM (Eq.~\ref{eq:C12}). As the latter quantity is not available within the 2-RDM propagation, reconstruction by a functional $D^{\rm R}_{123}$ depending on $D_{12}$ is required. We first test the performance of the reconstruction functionals $D^{\rm R}_{123}$, specifically the Valdemoro functional $D^{\rm V}_{123}$ (Eq.~\ref{eq:val}) and the contraction-consistent functional $D^{\rm C}_{123}$ introduced in Sec.~\ref{sec:con}. In the first step, we employ as input to these functionals the exact $D^{\rm exact}_{12}$ generated from the simultaneous propagation of $\Psi^{\rm MCTDHF}(t)$. Using the resulting $D^{\rm V}_{123}[D^{\rm exact}_{12}]$ and $D^{\rm C}_{123}[D^{\rm exact}_{12}]$ in the propagation of $D_{12}$ allows the assessment of the accuracy of the functionals decoupled from the error in $D_{12}$ accumulated during the propagation. We find excellent agreement for the dipole moment when the collision operator $C_{12}$ in Eq.~\ref{eq:C12} is calculated from the reconstructed $D^{\rm V}_{123}[D^{\rm exact}_{12}]$ and $D^{\rm C}_{123}[D^{\rm exact}_{12}]$ (Fig.~\ref{fig:first_level}). For the latter the agreement is clearly better. The error in the reconstructed collision operator $C^{\rm R}_{12}$ as determined by the basis independent measure 
\begin{align}
\epsilon^2=\text{Tr}_{12}[(C_{12}-C^{\rm R}_{12})^2]
\end{align}
is more than nine times smaller for $D_{123}^{\mathrm C}$ than for $D_{123}^{\mathrm V}$ (not shown). Despite the good agreement for the dipole moment we observe that the $N$-representability is not conserved during propagation: at $\tau\approx0.1$ the lowest eigenvalue (i.e.~geminal occupation number) of the 2-RDM drops to $g_{\text{min}}\approx -0.02$ for $D_{123}^{\mathrm C}$ and to $g_{\text{min}}\approx -0.05$ for $D_{123}^{\mathrm V}$ before it starts to oscillate keeping the eigenvalues bounded from below. These oscillations can be found in other 2-positivity conditions as well. The appearence of a lower bound for $g_{\text{min}}$ allows for a stable propagation of $D_{12}$ shown in Fig.~\ref{fig:first_level}. 
 
\subsubsection{Sensitivity of the reconstruction functional to errors in $D_{12}$}\label{sec:sen}

\begin{figure}
\centering
\includegraphics[width=\linewidth]{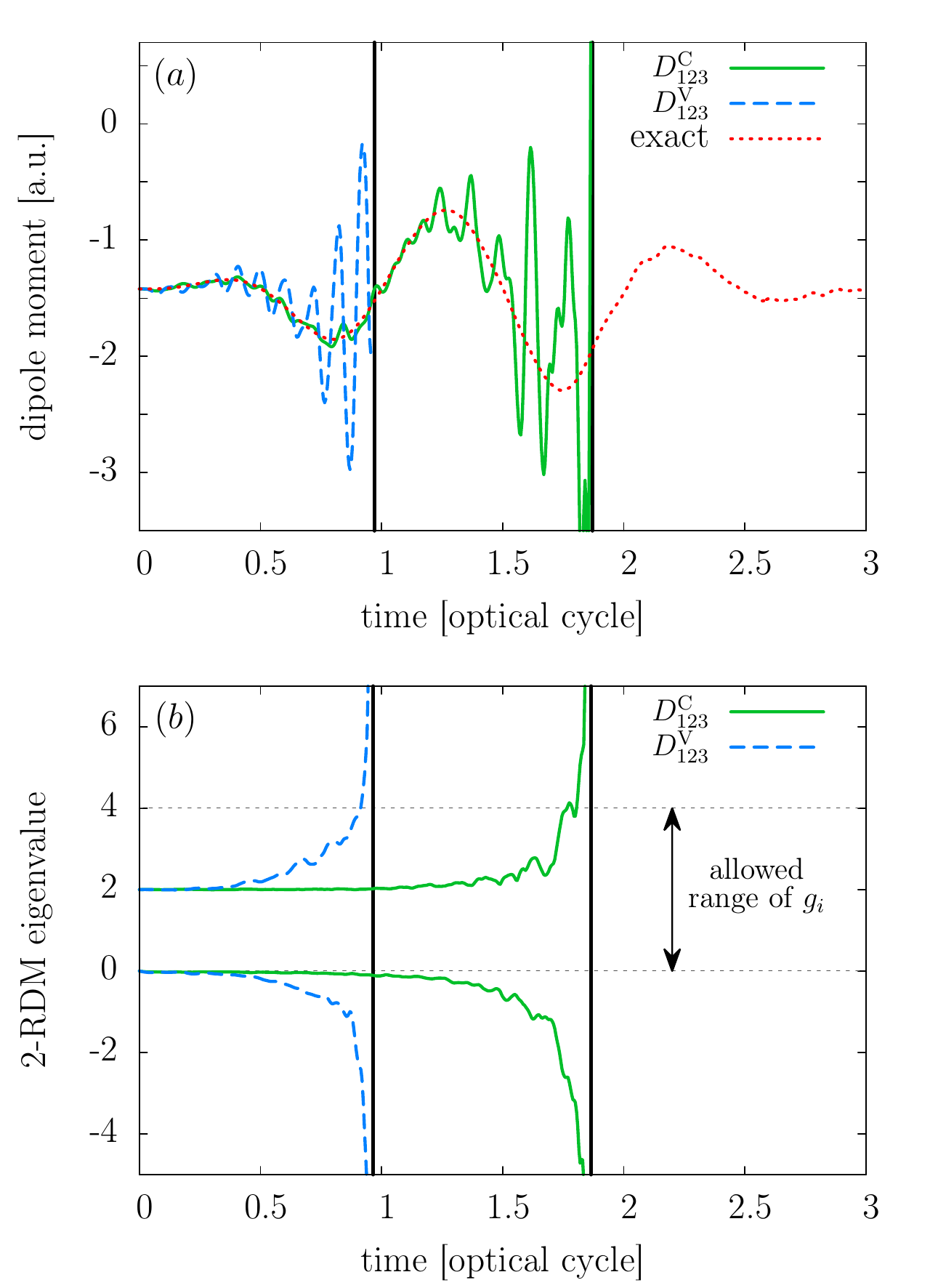}
\caption{(Color online) (a) Dipole moment of LiH subject to the laser pulse (Fig.~\ref{fig:pulse_form}) with $I=10^{14}\rm{W/cm^2}$ 
employing the reconstruction functionals $D_{123}^{\mathrm V}[D_{12}]$ and $D_{123}^{\mathrm C}[D_{12}]$ with $D_{12}$ propagated by Eq.~\ref{eq:2p_eom_orbital} while the orbitals are calculated via Eq.~\ref{eq:dphi_MCTDH} using $D^{\rm exact}_{12}$ from a parallel MCTDHF calculation. The violation of $N$-representability leads to divergence of the dipole moment. The point of divergence is marked by vertical lines for each reconstruction. Note that contraction consistency postpones but does not prevent the divergence. (b) The smallest and largest eigenvalue of the 2-RDM. The violation of $N$-representability clearly visible from eigenvalues outside the allowed range $0 \le g_i \le 4$ (marked by dashed horizontal lines) occurs in close temporal proximity to the divergence of the dipole moment.}
\label{fig:second_level_no_pu}
\end{figure}

\begin{figure}
\centering
\includegraphics[width=\linewidth]{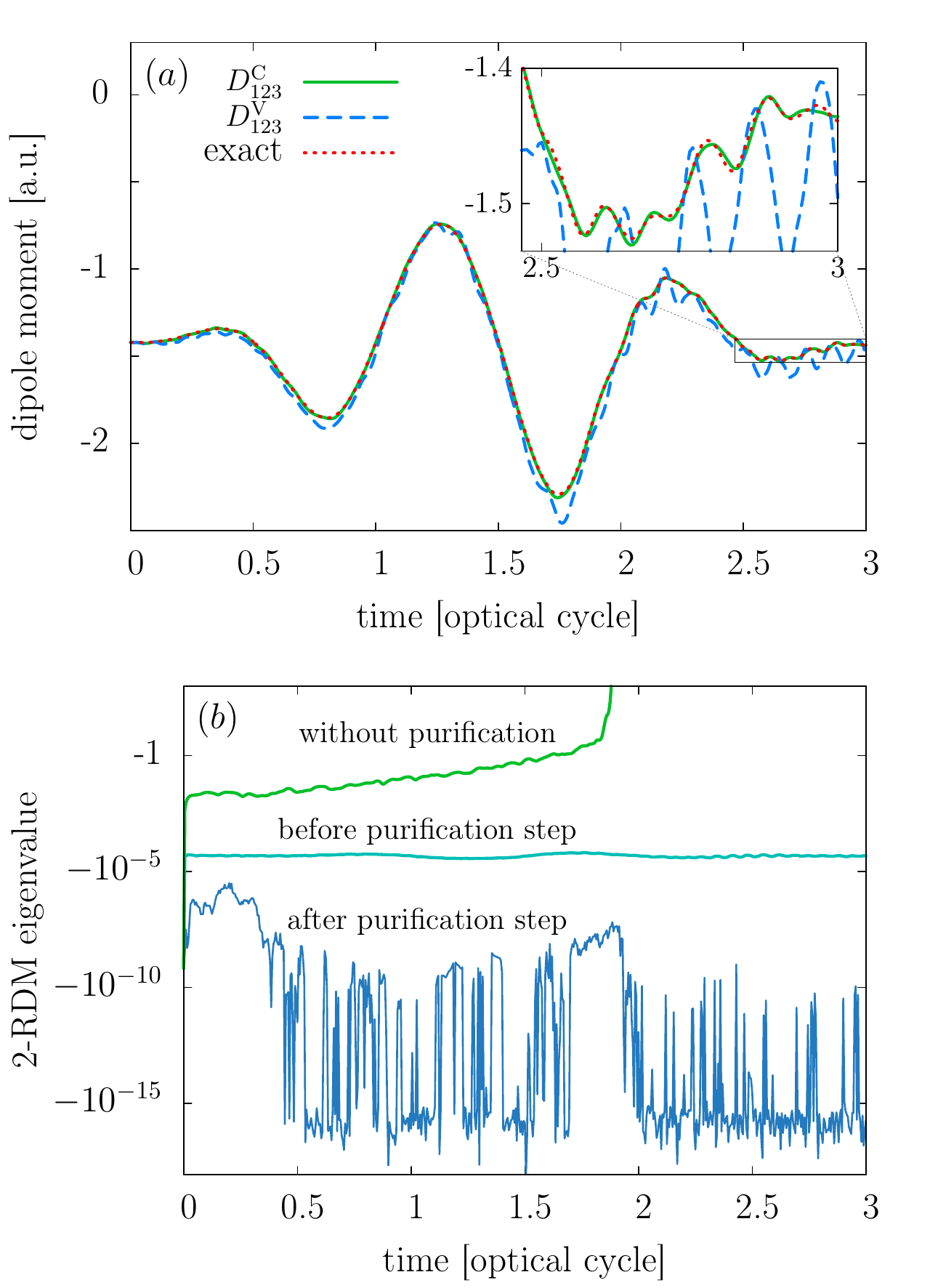}
\caption{(Color online) (a) Dipole moment for the same parameters as in Fig.~\ref{fig:second_level_no_pu} but with dynamical purification. (b) The smallest eigenvalue of the 2-RDM for the propagation employing $D_{123}^{\mathrm C}$ without purification [green line, compare Fig.~\ref{fig:second_level_no_pu}(b)] and with purification. For the latter the smallest eigenvalue is shown before and after each dynamical purification step (Eq.~\ref{eq:projection}). The negative occupation number after purification with 10 iterations is in general smaller than $10^{-9}$ indicating rapid convergence.}
\label{fig:second_level_with_pu}
\end{figure}

Successively approaching a realistic simulation scenario, we now test the sensitivity of reconstruction functionals $D^{\rm R}_{123}[D_{12}]$ to errors in $D_{12}$ when the exact 2-RDM is not available. On this level of approximation the collision operator $C^{\rm R}_{12}[D_{12}]$ induces a highly nonlinear feedback loop that tends to rapidly magnify the errors of $D_{12}$ accumulated during the evolution. At this stage, we still employ the exact $D_{12}^{\rm exact}$ in the propagation of the orbitals (Eq.~\ref{eq:dphi_MCTDH}) in order to decouple the error accumulation through the nonlinear orbital equation of motion from that of the nonlinear equation of motion for $D_{12}$ (Eq.~\ref{eq:2p_eom_orbital}). Focusing for the moment only on the latter, this nonlinear feedback loop ultimately produces severe instabilities such that the 2-positivity conditions of the 2-RDM are strongly violated causing, in turn, the divergence in physical observables such as the dipole moment (see Fig.~\ref{fig:second_level_no_pu}). This instability is present when using either $D_{123}^{\rm V}$ or $D_{123}^{\rm C}$. However, the onset of the divergence is delayed for the latter [see Fig.~\ref{fig:second_level_no_pu} (a)] providing an additional indication that an accurate reconstruction functional is key for the reliable propagation over finite times. The violation of the 2-RDM positivity conditions in terms of excursions of the eigenvalues outside of the allowed range, $0 \le g_i \le N$, shows that due to the nonlinear error magnification $N$-representability will not be preserved during propagation unless purification after each time step is enforced.
\subsubsection{Test of dynamical purification}

\begin{figure}[t]
\centering
\includegraphics[width=\linewidth]{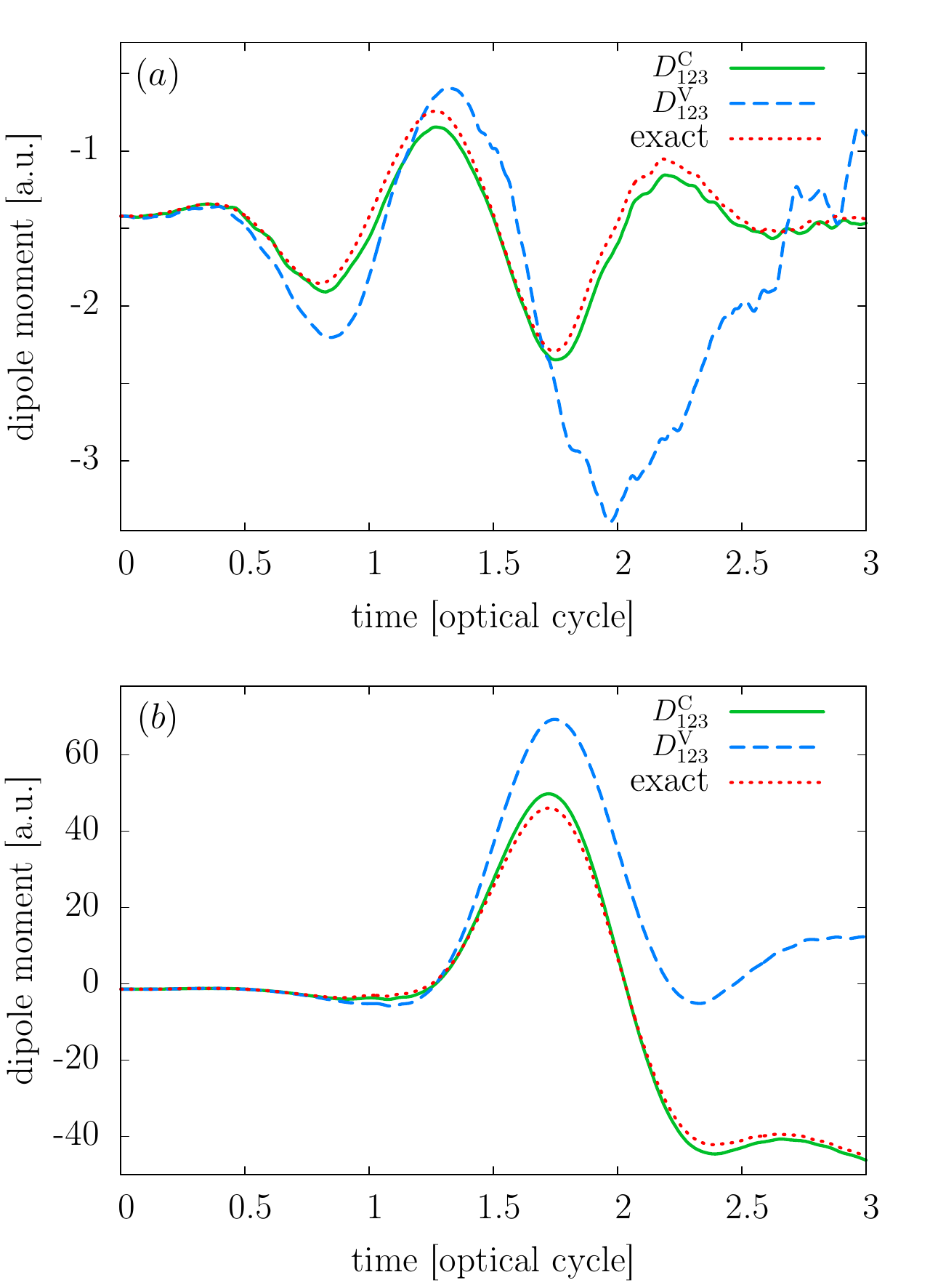}
\caption{(Color online) Dipole moment of LiH subject to the laser pulse (Fig.~\ref{fig:pulse_form}) with (a) $I=10^{14}\rm{W/cm^2}$ and (b) $I=8\times 10^{14}\rm{W/cm^2}$ employing the reconstruction functionals $D_{123}^{\mathrm V}[D_{12}]$ and $D_{123}^{\mathrm C}[D_{12}]$ within a fully self-consistent propagation of the TD-2RDM compared with the exact (MCTDHF) reference.}
\label{fig:dipole_3level}
\end{figure}

\begin{figure}
\centering
\includegraphics[width=\linewidth]{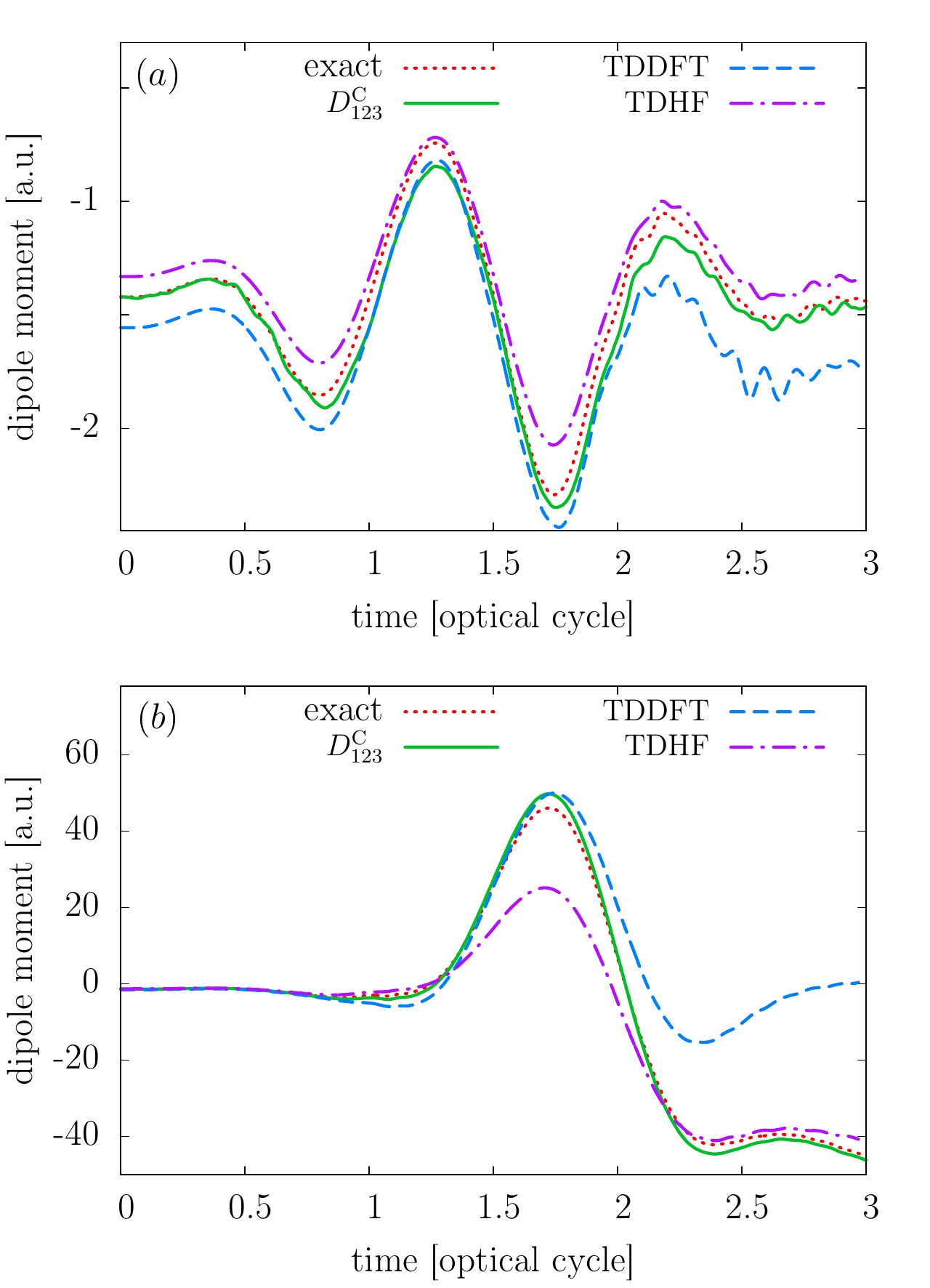}
\caption{(Color online) Dipole moment of LiH subject to the laser pulse (Fig.~\ref{fig:pulse_form}) with (a) $I=10^{14}\rm{W/cm^2}$ and (b) $I=8\times 10^{14}\rm{W/cm^2}$ for a fully self-consistent propagation of the TD-2RDM method compared with the exact (MCTDHF) reference, the TDHF and the TDDFT calculations. }
\label{fig:LiH_TDDFT}
\end{figure}

To preserve $N$-representability and to achieve a stable propagation of $D_{12}$, implementation of dynamical purification is essential. For each time step, Eq.~\ref{eq:purif_corr} - Eq.~\ref{eq:pur_2RDM2} adapted to spin symmetry are iteratively solved. Convergence is typically reached after $10$ iterations where the magnitude of the lowest eigenvalue of the 2-RDM and the 2-HRDM is reduced to values below $10^{-9}$ [see Fig.~\ref{fig:second_level_with_pu} (b)]. Repeating the test for sensitivity to propagation errors in $D_{12}$ including now the dynamical purification we find reasonable agreement for the time evolution of the dipole moment with $D_{123}^{\rm V}$ and perfect agreement with $D_{123}^{\rm C}$ relative to the MCTDHF reference [see Fig.~\ref{fig:second_level_with_pu} (a)]. The small amplitude oscillations after the conclusion of the pulse signifying the superposition of the ground state and excited states are reproduced with high accuracy while they are overestimated by $D_{123}^{\rm V}$ [see Fig.~\ref{fig:second_level_with_pu} (a)]. The correction of the error accumulation for $D_{12}$ by the purification thus dramatically improves the stability and accuracy of the propagation.
\subsubsection{Self-consistent propagation}

A fully self-consistent propagation requires one additional step: the use of the approximate $D_{12}$ also as input for the orbital equations of motion (Eq.~\ref{eq:dphi_MCTDH}). Up to this point we have used $D_{12}^{\rm exact}$ in Eq.~\ref{eq:dphi_MCTDH} in order to disentangle the error occurring in the propagation of $D_{12}$ from that of the single-particle orbitals. Since the latter is also a system of nonlinear equations containing both $D_{12}$ and the inverse of $D_1$ error magnification is to be expected. This (up to exponential) error magnification imposes an additional constraint on the required accuracy of the reconstruction as well as purification. Indeed, within a fully self-consistent propagation the simple reconstruction $D_{123}^{\rm V}$ is not able to accurately reproduce the time evolution of the dipole moment (see Fig.~\ref{fig:dipole_3level}). Strong deviations from the exact result occur due to the violation of the spin symmetries discussed in Sec.~\ref{sec:def}. It turns out that the conservation of these symmetries by $D_{123}^{\rm C}$ is essential to obtain results that are in agreement with the MCTDHF calculation (see Fig.~\ref{fig:dipole_3level}).\\
We also compare the present results for $d(t)$ with the prediction by the TDHF method and TDDFT within the adiabatic local density approximation (LDA) (for details see Appendix \ref{sec:tddft}). Within TDDFT and TDHF we use the corresponding DFT and HF ground states as the initial states which leads to the discrepancy at $t=0$ in the dipole moment. While for the lower intensity, this discrepancy is only moderately increased during the evolution, for the higher intensity dramatic enhancement of these deviations is observed. By contrast, the present TD-2RDM method performs consistently better than the TDHF and the TDDFT calculations over the entire time interval for both intensities [see Fig.~\ref{fig:LiH_TDDFT} (a) and (b)]. We note that the TDHF calculation can be viewed as a special case of the self-consistent propagation of the 2-RDM when the number of spin orbitals equals the number of electrons. In this case the reconstruction is exact and purification is not necessary since the resulting 2-RDM is $N$-representable at all times. While the TDHF and the TDDFT methods feature stability, they do not achieve accuracy [see Fig.~\ref{fig:LiH_TDDFT} (a) and (b)]. \\

\begin{figure}
\centering
\includegraphics[width=\linewidth]{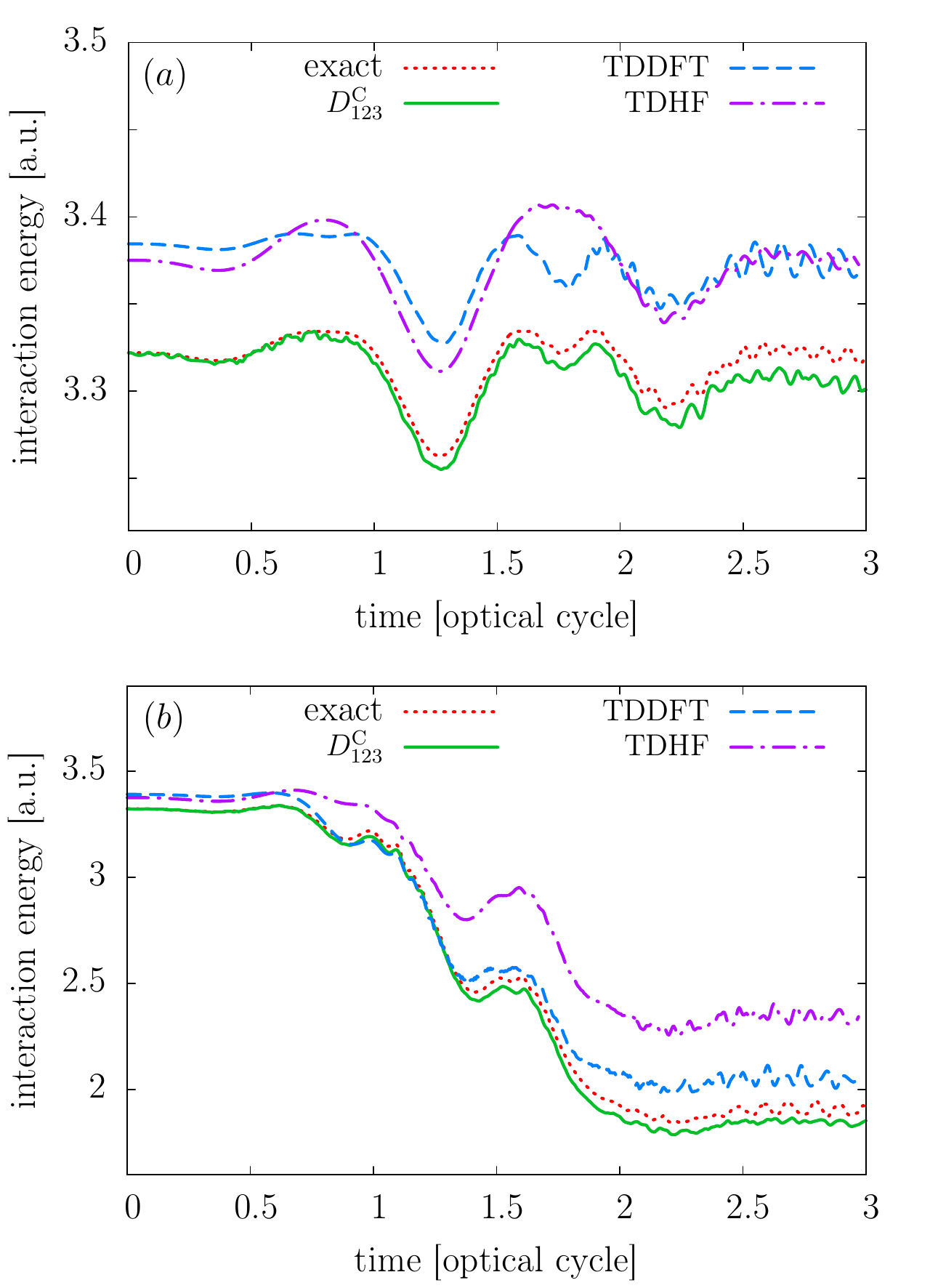}
\caption{(Color online) Time-dependent electron-electron interaction energy of LiH subject to the laser pulse (Fig.~\ref{fig:pulse_form}) with $I=10^{14}\rm{W/cm^2}$ (a) and $I=8\times 10^{14}\rm{W/cm^2}$ (b) for a fully self-consistent propagation of the TD-2RDM method compared with the exact (MCTDHF) reference, the TDHF and the TDDFT calculations.}
\label{fig:LiH_TDDFT_energy}
\end{figure}

\begin{figure}
\centering
\includegraphics[width=\linewidth]{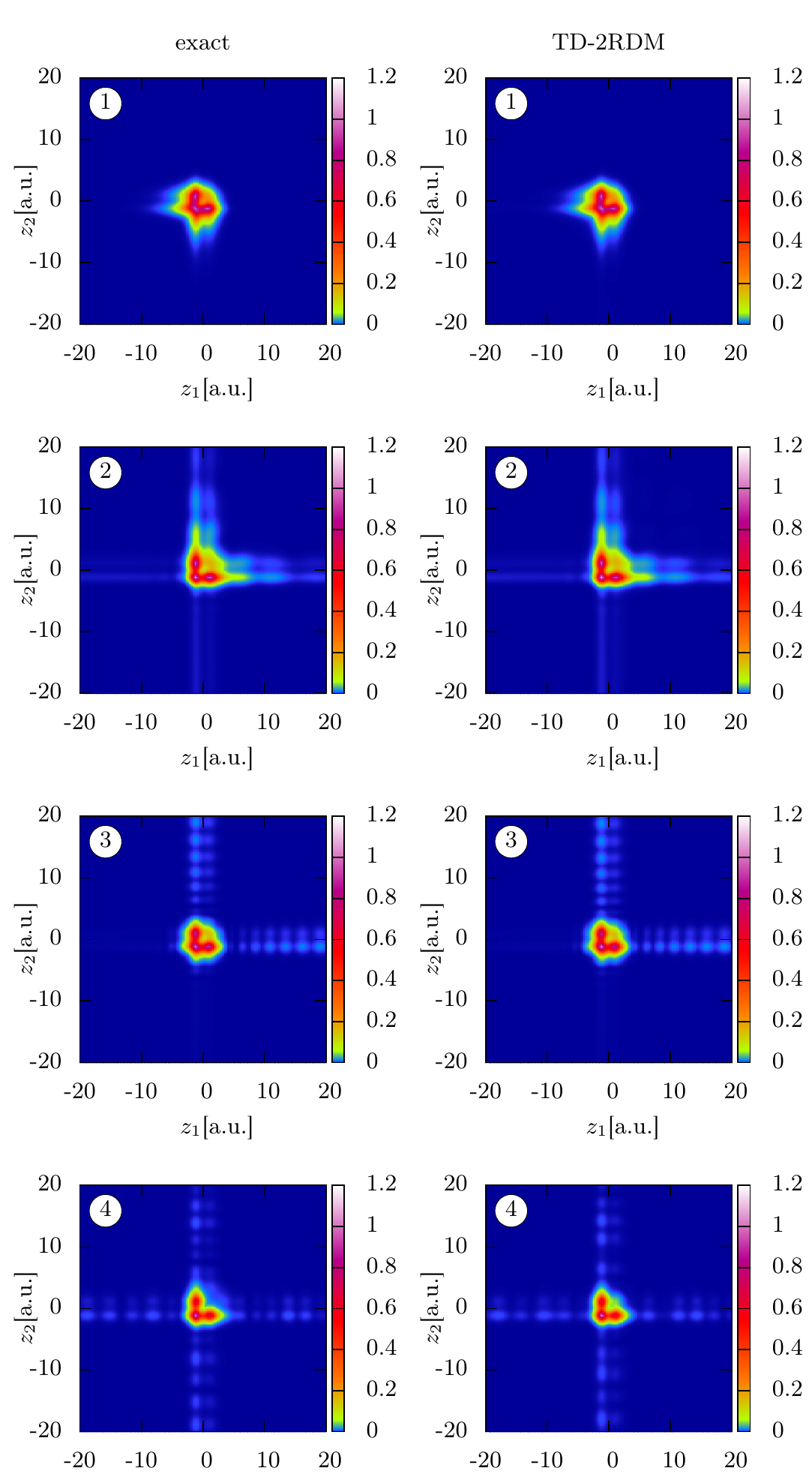}
\caption{(Color online) Pair density of the LiH molecule in the strong laser pulse with $I=8\times10^{14}\rm{W/cm^2}$ left column exact (MCTDHF); right column self-consistent propagation of the 2-RDM using the reconstruction function $D_{123}^{\mathrm C}[D_{12}]$. The pair density is shown at four times depicted in Fig.~\ref{fig:pulse_form} [rows (1) to (4)]. The stretched-out arms in the pair density are signatures of single particle ionization. The approximate distributions are in very good agreement with the exact result. Small differences appear at times (3) and (4).}
\label{fig:pair_density}
\end{figure}
\subsubsection{Two-particle observables}
A more stringent benchmark for the accuracy of the TD-2RDM method are two-particle observables. Unlike one-particle observables such as the dipole moment, calculation of these represents a major challenge as, in general, unknown or poorly known extraction functionals for mean-field descriptions have to be invoked to determine two-particle expectation values from the time-evolved density $\rho(z,t)$.\\
As an example we consider the two-particle interaction energy (Eq.~\ref{eq:E_int}). The present calculation for LiH [Fig.~\ref{fig:LiH_TDDFT} (a) and (b)] shows that the self-consistent TD-2RDM with $D_{123}^{\rm C}$ yields excellent agreement with the exact result and thus accounts for almost $100\%$ of the interaction energy unlike TDDFT or the TDHF method.
The time evolution of the interaction energy of LiH for high laser intensity $I=8\times 10^{14} W$ shows clear signatures of ionization [Fig.~\ref{fig:LiH_TDDFT} (b)]. The regions with steep reduction correspond to time intervals where the electron emission from the molecule preferably occurs. This leads to decrease of electron density and of interaction energy. At the plateaus the field reverses its sign and the electron density stays nearly constant before ionization occurs into the opposite direction. The small increase in interaction energy around $\tau\approx1.5$ indicates that the ionized electron is re-scattered at the molecule. \\
The spatio-temporal variation of the ionization process becomes directly visible in the pair density $\rho(z_1,z_2,t)$. The snapshots (Fig.~\ref{fig:pair_density}) at different times (marked in Fig.~\ref{fig:pulse_form}) display the pair density near the ground state (1), near the field maximum (2), at the time of re-scattering (3), and near the conclusion of the pulse (4). Overall, the agreement between the exact pair density and the one calculated by the TD-2RDM is excellent and differences are hardly visible. Minor deviations appear only after the re-scattering of electrons near $\tau\approx1.5$ close to time (3). At this time the electron-electron scattering rate is slightly underestimated since the approximation of at most two simultaneously interacting particles underlying the TD-2RDM description is less accurate. The approximated pair distribution is again in almost perfect agreement with the exact MCTDHF calculation after $\tau\approx1.5$. The fact that ionization happens almost exclusively along the coordinate axes with $z_1\approx0$ or $z_2\approx0$ shows that single ionization is the dominant contribution and double ionization which would show up along the diagonals $|z_1|=|z_2|$ is comparatively weak at this field strength.

\section{Conclusions and outlook}
We have presented a promising time-dependent many-body theory with polynomial scaling in particle number. The theory is based on the propagation of the time-dependent two-particle reduced density matrix (TD-2RDM) without invoking the $N$-particle wavefunction. One key ingredient is the reconstruction of the 3-RDM via the 2-RDM, a prerequisite for closing the equation of motion. We have presented a new reconstruction functional for the three particle reduced density matrix (3-RDM) which guarantees conservation of norm, energy, and spin during time propagation. In the reconstruction functional we have included those parts of the three-particle cumulant that can be reconstructed from the 2-RDM. For achieving stable propagation, a second key ingredient is crucial: due to the nonlinearity of the equation of motion, small errors rapidly (up to exponentially) magnify destroying $N$-representability. We have therefore devised a dynamical purification protocol that iteratively restores $N$-representability after each time step by enforcing the positivity of the 2-RDM and the two-hole reduced density matrix (2-HRDM). As a benchmark calculation we have applied the TD-2RDM method to the dynamics of electrons in the one-dimensional LiH molecule in strong laser fields and have compared the results to that of the time-dependent Hartree-Fock (TDHF) method, the full multiconfigurational time-dependent Hartree-Fock method (MCTDHF), and time-dependent density functional theory (TDDFT). We observe that the TD-2RDM method shows very good agreement with the MCTDHF results. The latter have been carefully checked for convergence and can serve as representative of the numerically exact four-electron wavefunction of this problem. As test observables we have used the dipole moment as a bona fide one-particle observable of great importance for the (non-)linear response to strong laser fields, and the electron-electron interaction energy and pair density as generic two-particle observables. For two-particle observables the 2-RDM method features the decisive advantage over effective one-particle descriptions such as TDDFT that the observable is directly accessible without invoking any read-out functionals.\\
We anticipate that the present TD-2RDM theory should provide a tool to accurately describe a wide variety of many-body systems as long as the dynamics is given by a sequence of two-particle interactions. Genuine three-particle correlations are neglected in our theory. Applications to other systems and larger numbers of degrees of freedom are envisioned. 
\section*{Acknowledgments}
We thank Ofir Alon for helpful discussions on multiconfigurational methods, Kazuhiro Yabana for hints on literature, and Florian Libisch and Georg Wachter for their help with TDDFT calculations. This work has been supported by the FWF doctoral school Solids4Fun, FWF SFB-041 ViCoM, and FWF SFB-049 Next Lite, and in part by KAKENHI (No.~23750007, No.~23656043, No.~23104708, No.~25286064, No.~26390076, and No.~26600111), the Photon Frontier Network Program of MEXT (Japan), and Center of Innovation Program from JST (Japan). Calculations have been performed on the Vienna Scientific Cluster 1.
\appendix
\section{The unitary decomposition of hermitian three-particle matrices with arbitrary symmetry}\label{sec:uni}

The unitary decomposition of $p$-particle matrices \cite{ColAbs80,CasMarHar86,ChiGuo83,SunLiTan84_2} is the generalization of the unitary decomposition of two-particle matrices which has been developed for hermitian antisymmetric two-particle matrices\footnote{As a side remark we note that the unitary decomposition of two-particle matrices is equivalent to the Ricci decomposition of general relativity which is used to define the trace free part of the Riemann curvature tensor known as the Weyl tensor \cite{Wal87}.} (for a review see, e.g., \cite{Maz07}). Briefly, any hermitian antisymmetric two-particle matrix $M_{12}$ can be decomposed into
\begin{align}\label{eq:decom_2RDM}
M_{12}=M_{12;\perp} + M_{12;\text{K}},
\end{align}	
where $M_{12;\text{K}}$ is the contraction-free component lying in the kernel of the contraction operator
\begin{align}
	\text{Tr}_2\left(M_{12;\text{K}}\right)=0,
\end{align}
and
\begin{align}\label{eq:ansatz_d12}
	M_{12;\perp}&= \frac{4}{r-2}M_1 \wedge I-\frac{4\text{Tr}_{1}(M_{1})}{(r-1)(r-2)}I \wedge I,
\end{align}
is an element of the orthogonal complement with $M_1=\mathrm{Tr}_{2}M_{12}$, $I$ is the identity and $r$ the number of orbitals. The component $M_{12;\perp}$ is orthogonal to the contraction-free component $M_{12;\text{K}}$ with respect to the Frobenius inner product for matrices \cite{Har02}
\begin{align}
	\text{Tr}_{12} \left( M_{12;\perp}M_{12;\text{K}} \right)=0.
\end{align}
Similar to Eq.~\ref{eq:ansatz_d12}, the unitary decomposition for hermitian symmetric two-particle matrices reads \cite{SunLiTan84}
\begin{align}\label{eq:ansatz_d12_sym}
M_{12;\perp}&= \frac{4}{r+2}M_1 \odot  I-\frac{4\text{Tr}_{1}(M_{1})}{(r+1)(r+2)}I\odot I,
\end{align}
where the symmetric product $\odot$ is defined in analogy to the antisymmetric wedge product $\wedge$ (Eq.~\ref{eq:wedge}). Note that the orthogonal component $M_{12;\perp}$ defined in Eq.~\ref{eq:ansatz_d12} and Eq.~\ref{eq:ansatz_d12_sym} depends only on the contraction of the two-particle matrix $M_1=\mathrm{Tr}_{2}M_{12}$. For the unitary decomposition of hermitian two-particle matrices with arbitrary symmetry \cite{alc04} the orthogonal component $M_{12;\perp}$ is uniquely determined from all diagonal and off-diagonal contractions of the two-particle matrix $M_{12}$. \\
We extend now this unitary decomposition to hermitian three-particle matrices $M_{123}$ with arbitrary symmetry
\begin{align}\label{eq:unit_decom}
M_{123}=M_{123;\perp}+M_{123;\text{K}}.
\end{align}
In this decomposition $M_{123;\text{K}}$ is the contraction-free component in the kernel of the contraction operator
\begin{align}\label{eq:kernel}
L_{3}(M_{123;\text{K}})=0,
\end{align}
where $L_{3}$ denotes all diagonal and off-diagonal contractions. As we show below the orthogonal component $M_{123;\perp}$ can be written as a functional of the 9 one-fold contractions
\begin{align}\label{eq:one-fold_contr}
&{}^1M^{i_1i_2}_{j_1j_2}=\sum_k M^{i_1i_2k}_{j_1j_2k} 
\qquad{}^2M^{i_1i_2}_{j_1j_2}=\sum_k M^{i_1i_2k}_{j_1kj_2} \nonumber\\
&{}^3M^{i_1i_2}_{j_1j_2}=\sum_k M^{i_1i_2k}_{kj_1j_2} 
\qquad{}^4M^{i_1i_2}_{j_1j_2}=\sum_k M^{i_1ki_2}_{j_1j_2k} \nonumber\\
&{}^5M^{i_1i_2}_{j_1j_2}=\sum_k M^{i_1ki_2}_{j_1kj_2} 
\qquad{}^6M^{i_1i_2}_{j_1j_2}=\sum_k M^{i_1ki_2}_{kj_1j_2} \nonumber\\
&{}^7M^{i_1i_2}_{j_1j_2}=\sum_k M^{ki_1i_2}_{j_1j_2k}
\qquad{}^8M^{i_1i_2}_{j_1j_2}=\sum_k M^{ki_1i_2}_{j_1kj_2} \nonumber\\
&{}^9M^{i_1i_2}_{j_1j_2}=\sum_k M^{ki_1i_2}_{kj_1j_2},
\end{align}
the 18 two-fold contractions
\begin{align}\label{eq:two-fold_contr}
{}^1M^{i}_{j}&=     \sum_{k_1k_2} M^{ik_1k_2}_{jk_1k_2} 
\qquad{}^2M^{i}_{j}=\sum_{k_1k_2} M^{ik_1k_2}_{k_1jk_2}  \nonumber\\
{}^3M^{i}_{j}&=     \sum_{k_1k_2} M^{ik_1k_2}_{k_1k_2j}  
\qquad{}^4M^{i}_{j}=\sum_{k_1k_2} M^{ik_1k_2}_{jk_2k_1} \nonumber\\
{}^5M^{i}_{j}&=     \sum_{k_1k_2} M^{ik_1k_2}_{k_2jk_1}  
\qquad{}^6M^{i}_{j}=\sum_{k_1k_2} M^{ik_1k_2}_{k_2k_1j}  \nonumber\\
{}^7M^{i}_{j}&=     \sum_{k_1k_2} M^{k_1ik_2}_{jk_1k_2} 
\qquad{}^8M^{i}_{j}=\sum_{k_1k_2} M^{k_1ik_2}_{k_1jk_2}  \nonumber\\
{}^9M^{i}_{j}&=     \sum_{k_1k_2} M^{k_1ik_2}_{k_1k_2j}  
\qquad{}^{10}M^{i}_{j}=\sum_{k_1k_2} M^{k_1ik_2}_{jk_2k_1} \nonumber\\
{}^{11}M^{i}_{j}&=     \sum_{k_1k_2} M^{k_1ik_2}_{k_2jk_1}  
\qquad{}^{12}M^{i}_{j}=\sum_{k_1k_2} M^{k_1ik_2}_{k_2k_1j}  \nonumber\\
{}^{13}M^{i}_{j}&=     \sum_{k_1k_2} M^{k_1k_2i}_{jk_1k_2} 
\qquad{}^{14}M^{i}_{j}=\sum_{k_1k_2} M^{k_1k_2i}_{k_1jk_2}  \nonumber\\
{}^{15}M^{i}_{j}&=     \sum_{k_1k_2} M^{k_1k_2i}_{k_1k_2j}  
\qquad{}^{16}M^{i}_{j}=\sum_{k_1k_2} M^{k_1k_2i}_{jk_2k_1} \nonumber\\
{}^{17}M^{i}_{j}&=     \sum_{k_1k_2} M^{k_1k_2i}_{k_2jk_1}  
\qquad{}^{18}M^{i}_{j}=\sum_{k_1k_2} M^{k_1k_2i}_{k_2k_1j},  
\end{align}
and the 6 three-fold contractions 
\begin{align}\label{eq:three-fold_contr}
&{}^1M=\sum_{k_1k_2k_3} M^{k_1k_2k_3}_{k_1k_2k_3} 
\qquad{}^2M=\sum_{k_1k_2k_3} M^{k_1k_2k_3}_{k_1k_3k_2} \nonumber\\
&{}^3M=\sum_{k_1k_2k_3} M^{k_1k_2k_3}_{k_2k_1k_3}  
\qquad{}^4M=\sum_{k_1k_2k_3} M^{k_1k_2k_3}_{k_2k_3k_1} \nonumber\\
&{}^5M=\sum_{k_1k_2k_3} M^{k_1k_2k_3}_{k_3k_1k_2}  
\qquad{}^6M=\sum_{k_1k_2k_3} M^{k_1k_2k_3}_{k_3k_2k_1}.  
\end{align}
Generalizing the linear expansion for two-particle matrices \cite{alc04} we expand the orthogonal component of the three-particle matrix as
\begin{align}\label{eq:ansatz}
[M_{\perp}]^{i_1i_2i_3}_{j_1j_2j_3}=
&\sum_{k=1}^{6} \sum_{\tau \in S_3}
a^k_{\tau} \;
\delta^{i_{1}}_{j_{\tau(1)}}
\delta^{i_{2}}_{j_{\tau(2)}}
\delta^{i_{3}}_{j_{\tau(3)}} \; {}^kM \nonumber \\
+&\sum_{k=1}^{18} \sum_{\substack{\sigma,\tau \in S_3\\\sigma(1)<\sigma(2)}}
b^k_{\tau,\sigma} \;
\delta^{i_{\sigma(1)}}_{j_{\tau(1)}}
\delta^{i_{\sigma(2)}}_{j_{\tau(2)}}
\;{}^kM^{i_{\sigma(3)}}_{j_{\tau(3)}} \nonumber \\
+&\sum_{k=1}^{9} \sum_{\sigma,\tau \in S_3}
c^k_{\tau,\sigma} \;
\delta^{i_{\sigma(1)}}_{j_{\tau(1)}}
\;{}^kM^{i_{\sigma(2)}i_{\sigma(3)}}_{j_{\tau(2)}j_{\tau(3)}},
\end{align}
where $S_3$ denotes the permutation group of three elements. 
The restriction $\sigma(1)<\sigma(2)$ in the second term is necessary since the Kronecker deltas ($\delta$) can be commuted without creating a new coefficient. In this expansion there are $6 \times 3!$ coefficients $a^k_{\tau}$, $18 \times 3! \times 3!/2$ coefficients $b^k_{\tau,\sigma}$ and $9 \times 3! \times 3!$ coefficients $c^k_{\tau,\sigma}$ for which we will use the short hand notation $\vec{a},\vec{b}$ and $\vec{c}$. To determine the coefficients we insert the expansion (Eq.~\ref{eq:ansatz}) into Eqs.~\ref{eq:one-fold_contr}. Note that Eqs.~\ref{eq:two-fold_contr} and Eqs.~\ref{eq:three-fold_contr} do not give an additional set of conditions since they are implied by Eqs.~\ref{eq:one-fold_contr}. In general the result of a one-fold contraction of the expansion Eq.~\ref{eq:ansatz} has the following form ($n\in\{1 \dots 9\}$)
\begin{align}\label{eq:condition}
{}^nM^{i_1i_2}_{j_1j_2}=
&\sum_{k=1}^{6} \sum_{\substack{ \mu,\nu \in S_2\\ \mu(1)<\mu(2)}}
f^{n,k}_{\mu,\nu}\big(\vec{a},\vec{b} \,\big) \;
\delta^{i_{\mu(1)}}_{j_{\nu(1)}}
\delta^{i_{\mu(2)}}_{j_{\nu(2)}} \; {}^kM \nonumber \\
+&\sum_{k=1}^{18} \sum_{\mu,\nu \in S_2}
h^{n,k}_{\mu,\nu}\big(\vec{b},\vec{c}\,\big) \;
\delta^{i_{\mu(1)}}_{j_{\nu(1)}}
{}^kM^{i_{\mu(2)}}_{j_{\nu(2)}} \nonumber \\
+&\sum_{k=1}^{9} \sum_{\mu,\nu \in S_2}
w^{n,k}_{\mu,\nu}\big(\vec{c}\,\big) \;
{}^kM^{i_{\mu(1)}i_{\mu(2)}}_{j_{\nu(1)}j_{\nu(2)}},
\end{align}
where $f^{n,k}_{\mu,\nu}\big(\vec{a},\vec{b} \big)$, $h^{n,k}_{\mu,\nu}\big(\vec{b},\vec{c}\big)$ and $w^{n,k}_{\mu,\nu}\big(\vec{c}\big)$ are linear functions of the coefficients, and $\mu,\nu$ are permutations in the permutation group $S_2$. In order for Eq.~\ref{eq:condition} to be an identity the terms on the right hand side containing either ${}^kM$ or ${}^kM^{i}_{j}$ must vanish and only the term containing ${}^nM^{i_1i_2}_{j_1j_2}$ appearing on the left hand side must remain. Consequently,
\begin{align}
f^{n,k}_{\mu,\nu}\big(\vec{a},\vec{b} \, \big)=0 \label{eq:first_cond}\\
h^{n,k}_{\mu,\nu}\big(\vec{b},\vec{c}\,\big)=0 \label{eq:second_cond}
\end{align}
and
\begin{align}
w^{n,k}_{\mu,\nu}\big(\vec{c}\,\big)&=0 \qquad \text{for}\; \mu,\nu \neq \text{id} \nonumber \\
w^{n,k}_{\text{id},\text{id}}\big(\vec{c}\,\big)&=\delta_{n}^k,
\label{eq:third_cond}
\end{align}
where id is the identity permutation. The $9\times3!\times3!$ coefficients $\vec{c}$ are uniquely determined by the $9\times9\times2!\times2!$ conditions given by Eq.~\ref{eq:third_cond}. Once the coefficients $\vec{c}$ are determined the $18\times3!\times3!/2$ coefficients $\vec{b}$ can be calculated from the $9\times18\times2!\times2!$ conditions of Eq.~\ref{eq:second_cond} and similarly the $6\times3!$ coefficients $\vec{a}$ can be calculated from the $9\times6\times2!\times2!/2$ conditions contained in Eq.~\ref{eq:first_cond}. Note that for the coefficients $\vec{a},\vec{b}$ there are more equations than variables so it is not a priori guaranteed that a solution exists. However, it turns out that the set of coupled equations Eq.~\ref{eq:first_cond}, Eq.~\ref{eq:second_cond}, and Eq.~\ref{eq:third_cond} has a unique solution for all orbital dimensions $r>4$. This shows that $M_{123;\perp}$ is a unique functional of the one-fold contractions. The solution for the coefficients depends solely on the number of orbitals $r$ since the only parameter that enters the equations is the trace of the Kronecker delta given by the orbital dimension $\sum_i\delta^i_i=r$. We solve the equations for the coefficients using symbolic computation performed with Mathematica. We find that all coefficients can be written in the following form
\begin{align}
X=&\frac{A_1}{r-4}+\frac{A_2}{r+4}+\frac{B_1}{r-3}+\frac{B_2}{r+3}\nonumber\\
+&\frac{C_1}{r-2}+\frac{C_2}{r+2}
+\frac{D_1}{r-1}+\frac{D_2}{r+1}+\frac{E_1}{r}+\frac{E_2}{r^2}
\end{align}
with rational coefficients $A_1, \dots ,E_2$. Obviously the coefficients are well defined only for $r>4$. A similar result holds also for the unitary decomposition of the 2-RDM for which $r>2$ has to be fulfilled. \\
Contrary to the case of arbitrary symmetry our application to the propagation of the 2-RDM requires the unitary decomposition of the $(\uparrow \uparrow \downarrow)$-block of the three-particle cumulant $\Delta_{123}^{\uparrow \uparrow \downarrow}$ which is antisymmetric in the first two indices. This significantly reduces the numerical effort to calculate the orthogonal part since in this case there are only 4 one-fold contractions, 5 two-fold contractions and 2 three-fold contractions and all other contractions can be expressed by these basic contractions. To explicitly evaluate $\Delta_{123;\perp}^{\uparrow \uparrow \downarrow}$ we calculate the basic contractions using $\Delta_{123}^{\uparrow \uparrow \downarrow}=D_{123}^{\uparrow \uparrow \downarrow}-D_{123}^{\mathrm{V}\uparrow \uparrow \downarrow}$ and insert them into Eq.~\ref{eq:ansatz} with the determined coefficients $\vec{a},\vec{b},\vec{c}$.

\section{TDDFT calculations in 1D}\label{sec:tddft}
While in three dimensions TDDFT is a well established theory to describe the dynamics of atomic, molecular, and solid state systems with a large number of electrons, one-dimensional TDDFT has been studied only very recently \cite{WagStoBur12,HelFukCas11}. The principal difference between one and three dimensions is that the Coulomb interaction~$\sim 1/\vert z-z' \vert$ leads to diverging interaction energies in 1D. This can be avoided by introducing the softened Coulomb interaction (Eq.~\ref{eq:w12}). The equations of motion in 1D of the time-dependent Kohn-Sham orbitals are \cite{Ull12}
\begin{align}
i \partial_t \phi^{\rm KS}_i(z,t) &= \Big(-\frac{1}{2}\frac{\partial^2}{\partial z^2}+V_{\rm eff}[\rho(z,t)] \Big) \phi^{\rm KS}_i(z,t),
\end{align}
with
\begin{align}
V_{\rm eff}[\rho(z,t)]=V_{\rm H}[\rho(z,t)]+V_{\rm x}[\rho(z,t)]+V_{\rm c}[\rho(z,t)],
\end{align}
where $V_{\rm H}[\rho]$ denotes the Hartree potential
\begin{align}
V_{\rm H}[\rho(z,t)]=\int \frac{\rho(z',t)}{\sqrt{(z-z')^2 +d}}\text{d}z',
\end{align}
$V_{\rm x}[\rho(z,t)]$ and $V_{\rm c}[\rho(z,t)]$ denotes the exchange and correlation potential, respectively. Within the local density approximation (LDA) the exchange and correlation potential is calculated from the uniform electron gas with the 1-RDM denoted by $D^{\rm unif}(z;z')$. The exchange potential $V_{\rm x}[\rho]$ for the 1D electron gas with softened Coulomb interaction can be evaluated analytically yielding a Meijer G-function \cite{WagStoBur12}
\begin{align}
V_{\rm x}[\rho]&=-\frac{1}{4}\frac{\delta}{\delta \rho}\int \frac{\vert D^{\rm unif}(z;z')\vert^2}{\sqrt{(z-z')^2 +d}}\text{d}z\text{d}z' \nonumber\\
&=-\frac{\rho}{4}G^{2,1}_{1,3} \!\left( \left. \begin{matrix} \frac{1}{2} \\ 0,0,-\frac{1}{2} \end{matrix} \; \right| \, d k_{\rm F}^2 \right),
\end{align} 
where $d$ is the Coulomb softening parameter and $k_{\rm F}=\frac{\pi \rho}{2}$. The correlation potential $V_{\rm c}[\rho]$ within LDA can be derived by quantum Monte Carlo calculations for the uniform 1D electron gas with soft Coulomb potential as discussed in \cite{HelFukCas11}.

%
\end{document}